\documentclass{aa}  
\usepackage{graphicx}
\usepackage{txfonts}
\begin{document} 
\title{  {\it Herschel}
\thanks{{\it Herschel} is an ESA space observatory with science instruments 
provided by European-led Principal Investigator consortia and with 
important participation from NASA.}
observations of extreme OH/IR stars}
\subtitle{the isotopic ratios of oxygen as a sign-post for the stellar mass}
\author{K. Justtanont\inst{\ref{inst1}}
\and M.J. Barlow\inst{\ref{ucl}}
\and J. Blommaert\inst{\ref{brussel},\ref{vito}}
\and L. Decin\inst{\ref{leuven}}
\and F. Kerschbaum\inst{\ref{vie}}
\and M. Matsuura\inst{\ref{ucl},\ref{cardiff}}
\and H. Olofsson\inst{\ref{inst1}}
\and P. Owen\inst{\ref{ucl}}
\and P. Royer\inst{\ref{leuven}}
\and B. Swinyard\inst{\ref{ucl},\ref{ral}}
\and D. Teyssier\inst{\ref{esac}}
\and L.B.F.M. Waters\inst{\ref{sron},\ref{UvA}}
\and J. Yates\inst{\ref{ucl}}
}
\institute{Chalmers University of Technology, Onsala Space Observatory,
   S-439 92 Onsala, Sweden\label{inst1}
\and University College London, Dept. of Physics \& Astronomy, Gower Street,
London, WC1E 6BT, UK\label{ucl}
\and Astronomy and Astrophysics Research Group, Dep. of Physics and 
Astrophysics, V.U. Brussel, Pleinlaan 2, 1050 Brussels, 
Belgium\label{brussel}
\and Flemish Institute of Technical Research, VITO, Mol, Belgium\label{vito}
\and Instituut voor Sterrenkunde, Katholieke Universiteit Leuven, 
Celestijnenlaan 200D, 3001 Leuven, Belgium\label{leuven}
\and University of Vienna, Department of Astrophysics, T\"urkenschanzstrasse 
17, 1180 Wien, Austria\label{vie}
\and School of Physics and Astronomy, Cardiff University, Queen's Buildings, 
The Parade, Cardiff, CF24 3AA, UK\label{cardiff}
\and Space Science and Technology Department, Rutherford Appleton Laboratory,
Oxfordshire OX11 0QX, UK\label{ral}
\and European Space Astronomy Centre, ESA, P.O. Box 78, E-28691
Villanueva de la Ca\~nada, Madrid, Spain\label{esac}
\and SRON Netherlands Institute for Space Research, Sorbonnelaan 2,
3584 CA Utrecht, The Netherlands\label{sron}
\and Sterrenkundig Instituut Anton Pannekoek, Universiteit van Amsterdam,
Postbus 94249, 1090 GE Amsterdam, The Netherlands\label{UvA}
}
\date{Received April 7, 2015; accepted May 11, 2015}

\abstract
{}
{
The late stages of stellar evolution are mainly governed by the mass
of the stars. Low- and intermediate-mass stars lose copious amounts of
mass during the asymptotic giant branch (AGB) which obscure the central star
making it difficult to study the stellar spectra and determine the stellar
mass. In this study, we present observational data that can be used
to determine lower limits to the stellar mass.
}
{
Spectra of nine heavily reddened AGB stars taken by the Herschel Space
Observatory display numerous molecular emission lines. The strongest
emission lines are due to H$_{2}$O. 
We search for the presence of isotopologues of H$_{2}$O in these objects.
}
{
We detected  the $^{16}$O and $^{17}$O isotopologues of water 
in these stars, but lines due to H$_{2}^{18}$O are absent.
The lack of $^{18}$O is predicted by a scenario where the star
has undergone  hot-bottom burning which preferentially destroys $^{18}$O
relative to $^{16}$O and $^{17}$O. From stellar evolution calculations,
this process is thought to occur
when the stellar mass is above 5~M$_{\odot}$ for solar metallicity.
Hence, observations of different isotopologues of H$_{2}$O can be used
to help determine the lower limit to the initial stellar mass.
}
{
From our observations, we deduce that these  extreme OH/IR stars are
intermediate-mass stars with masses of $\geq$ 5~M$_{\odot}$. Their high 
mass-loss rates of $\sim 10^{-4}$ M$_{\odot}$ yr$^{-1}$ may affect the
enrichment of the interstellar medium and the overall chemical evolution 
of our Galaxy.
}
{
\keywords{Stars: AGB and post-AGB -- Stars: evolution -- mass-loss -- 
Circumstellar matter -- Submillimeter: stars
}
\maketitle
%
\section{Introduction}
Low- and intermediate-mass stars (0.8-8 M$_{\odot}$) evolve onto the
asymptotic giant branch (AGB) after exhaustion of the central He. 
During this phase, the stars continue nucleosynthesis in a thin shell of
He, surrounded by a larger H-burning shell \citep[e.g.,][]
{iben83, habing96}. During its AGB lifetime, a star experiences a number
of He-flashes that lead to a sudden increase in its luminosity over
a brief period of time. 
Owing to efficient convection inside the star, the nucleosynthesis products
are mixed outwards (the so-called third dredge-up). Fresh carbon produced
in the He-shell burning is transported to the stellar photosphere and
will increase the carbon-to-oxygen ratio, which can turn the originally
oxygen-rich stars into carbon-rich stars.

The exact outcome of the nucleosynthesis and mixing events on the
stellar surface abundance depends on
the stellar mass. This is especially true when it comes to the process called
hot-bottom burning, where the base of the convective zone 
of the hydrogen envelope is hot
enough for CNO-cycle burning to destroy carbon \citep{lattanzio03}.
There is observational evidence that the process of hot-bottom burning occurs.
\cite{sackmann92}
observed a number of stars that exhibit anomalously high lithium abundance
which is thought to be a byproduct of hot-bottom burning. This
process is also thought to prevent a star from becoming carbon-rich.
The minimum limit of the stellar mass that can trigger this process is
estimated to be 
$\geq$ 5~M$_{\odot}$ for stars with a solar metallicity and can be smaller
for lower metallicities \citep{karakas14}. Another signpost that AGB stars
are descendants of such intermediate-mass main-sequence stars is 
a low $^{12}$C/$^{13}$C ratio -- during the AGB phase, more massive 
intermediate-mass stars enter the CNO cycle which tends to produce $^{13}$C
while converting $^{12}$C to $^{14}$N.
The presence of $^{13}$C is the main neutron source for the s-process elements
\citep{busso99, herwig05}. Another possible route is thought
to be from the $^{22}$Ne source in intermediate-mass stars, which requires a 
higher temperature.

An attempt to estimate masses of OH/IR stars near the galactic centre was
carried out by \cite{wood98}. They derived a mass of $\sim$ 4~M$_{\odot}$
for many of the stars in the sample 
using a period-luminosity relationship. 
Two stars with the longest periods were thought to have masses of up to 
7~M$_{\odot}$.
A number of optically visible OH/IR stars have been observed to exhibit
a high Li abundance but show weak s-process elements 
\citep{garcia07, garcia13}. These authors conclude that the stars have
undergone hot-bottom burning but that there is a mechanism that delays the
onset of s-process element production 
\citep{karakas12}. 
For OH/IR stars with an optically thick circumstellar envelope, 
which prevents the direct determination of stellar abundance,
other hot-bottom burning indicators must be used.
Two such stars (AFGL~5379 and OH~26.5+0.6) have been observed with
the Herschel Space Observatory \citep[hereafter {\it Herschel},][]{pilbratt10} 
to have strong water
emission lines in H$_{2}^{16}$O and H$_{2}^{17}$O but no detection of the
H$_{2}^{18}$O line \citep{justtanont13}. 
An AGB evolutionary model for a 5~M$_{\odot}$ star by \cite{lattanzio03}
shows that during hot-bottom burning, $^{18}$O is preferentially destroyed
with respect to the other two isotopes. The {\it Herschel} observations of
$^{18}$O/$^{17}$O ratios of $\ll$ 1 for these two stars 
are in contrast to the value of $\sim$ 3 derived from  
the interstellar medium \citep{wilson94}. 

In this paper, we present a larger sample of
extreme OH/IR stars -- those with very dusty circumstellar envelopes
such that silicate dust features at 10 and 20~$\mu$m are in absorption,
indicating high ($\dot{M} \geq 10^{-4}$~M$_{\odot}$ yr$^{-1}$) mass-loss 
rates. The observations of these stars were  
taken with all three instruments on board 
{\it Herschel} in order to search for
the emission lines of three isotopologues of H$_{2}$O as signposts
for hot-bottom burning, with the aim to obtain a lower limit to 
the stellar mass.
Detailed modelling of the line emission of H$_{2}$O and other detected 
molecules will be presented in the future
papers. In Sect. 2, we present the {\it Herschel} observations obtained
as part of an open-time program on OH/IR stars. We discuss the
results of our observations in Sect. 3 and summarize our findings
in Sect. 4.

\section{Observations}

We obtained {\it Herschel} spectra of eight OH/IR stars selected from
the sample based on \cite{justtanont06}. These stars all exhibit the 
silicate dust features in absorption at both 10 and 18$\mu$m and in some
cases also show a water-ice band at 3.1$\mu$m. Table~\ref{tab_all_sources}
lists all stars observed and presented in this paper 
for the first time using the three {\it Herschel} instruments PACS, 
HIFI, and SPIRE. As noted, some stars have had their spectra
taken as part of either a guaranteed time or another open-time program.

\begin{table*}
\caption{OH/IR stars observed in the present work with observation identifiers
(ObsID) indicated.}
\label{tab_all_sources}
\begin{tabular}{lllcccl}
\hline\hline
source &  RA (2000) & Dec (2000) & \multicolumn{3}{c}{ObsIDs} & note \\
\cline{4-6}
       &            &            & SPIRE & PACS & HIFI & \\
\hline
OH~127.8+0.0&  01 33 51.2 & +62 26 53.2 & 1342268319 & 1342189956-1342189961 & - & PACS cal\\
AFGL~5379   &  17 44 24.0 & -31 55 35.5 & 1342268287 & 1342228537/1342228538 & 1342250605 & 1,2\\
OH~21.5+0.5 &  18 28 31.6 & -09 58 10.7 & 1342268311 & 1342268748/1342268778 & - &  \\
OH~26.5+0.6 &  18 37 32.5 & -05 23 59.2 & 1342243624 & 1342207776/1342207777 & 1342244511 & 1,2,3\\
OH~30.7+0.4 &  18 45 53.1 & -01 46 58   & 1342268309 & 1342268789/1342268790 & 1342271263 & \\
OH~30.1-0.7 &  18 48 41.9 & -02 50 28.3 & 1342268316 & 1342269304/1342269305 & 1342271264 & \\
OH~32.0-0.5 &  18 51 26.2 & -01 03 52   & 1342268317 & 1342268791/1342268792 & 1342271265 & \\
OH~32.8-0.3 &  18 52 22.2 & -00 14 13.9 & 1342268318 & 1342268793/1342268794 & 1342271266 & \\
OH~42.3-0.1 &  19 09 07.5 & +08 16 22.5 & 1342268308 & 1342268797/1342268798 & - & \\
\hline
\end{tabular}
\tablefoot{The 
archived data taken from guaranteed time programs are indicated as
1 from MESS, 2 from HIFISTARS and 3 from another open-time program.
The PACS spectra of OH~127.8 were taken as part of the calibration time.}
\end{table*}

We observed four extreme OH/IR stars with the {\it Herschel}-HIFI instrument 
\citep{degraauw10}, which happened to be the last set of observations that
{\it Herschel} did before the helium ran out at the end of 
April 2013 (Fig.~\ref{fig_ddt}). The
frequency coverage for these stars are 1094.3-1098.4~GHz 
(lower side-band, LSB) and 1106.3-1110.4~GHz (upper side-band, USB).
The LSB frequency covers
the ortho-H$_{2}$O transition of 3$_{12}-3_{03}$ for all three
isotopologues while the USB permits the observations of the
ground state transition of para-H$_{2}^{17}$O 1$_{11}-0_{00}$.
The raw data (level 0) were processed with pipeline version SPG 11.0 
to obtain the level 1 and 2 data.
Further data reduction was performed in HIPE12\footnote{ 
HCSS / HSpot / HIPE is are joint developments by 
the Herschel Science Ground Segment Consortium, consisting of ESA, 
the NASA Herschel Science Center, and the HIFI, PACS, and SPIRE consortia.}
\citep{ott10} with calibration files
version HIFI\_CAL\_15. Both polarizations were averaged together and
the final spectra had the baseline subtracted and rebinned. From
Fig.~\ref{fig_ddt} it can bee seen that all objects, with the exception of 
OH~30.7+0.4, show the emission line due to H$_{2}^{16}$O. 
Using a routine in HIPE12, we converted the antenna temperature to a flux scale
so that we have the same flux scale for all three instruments, assuming a
point source.
The line fluxes are listed in Table~\ref{tab_obs_hifi}

\begin{figure}[h]
\centering
\resizebox{\hsize}{!}{\includegraphics{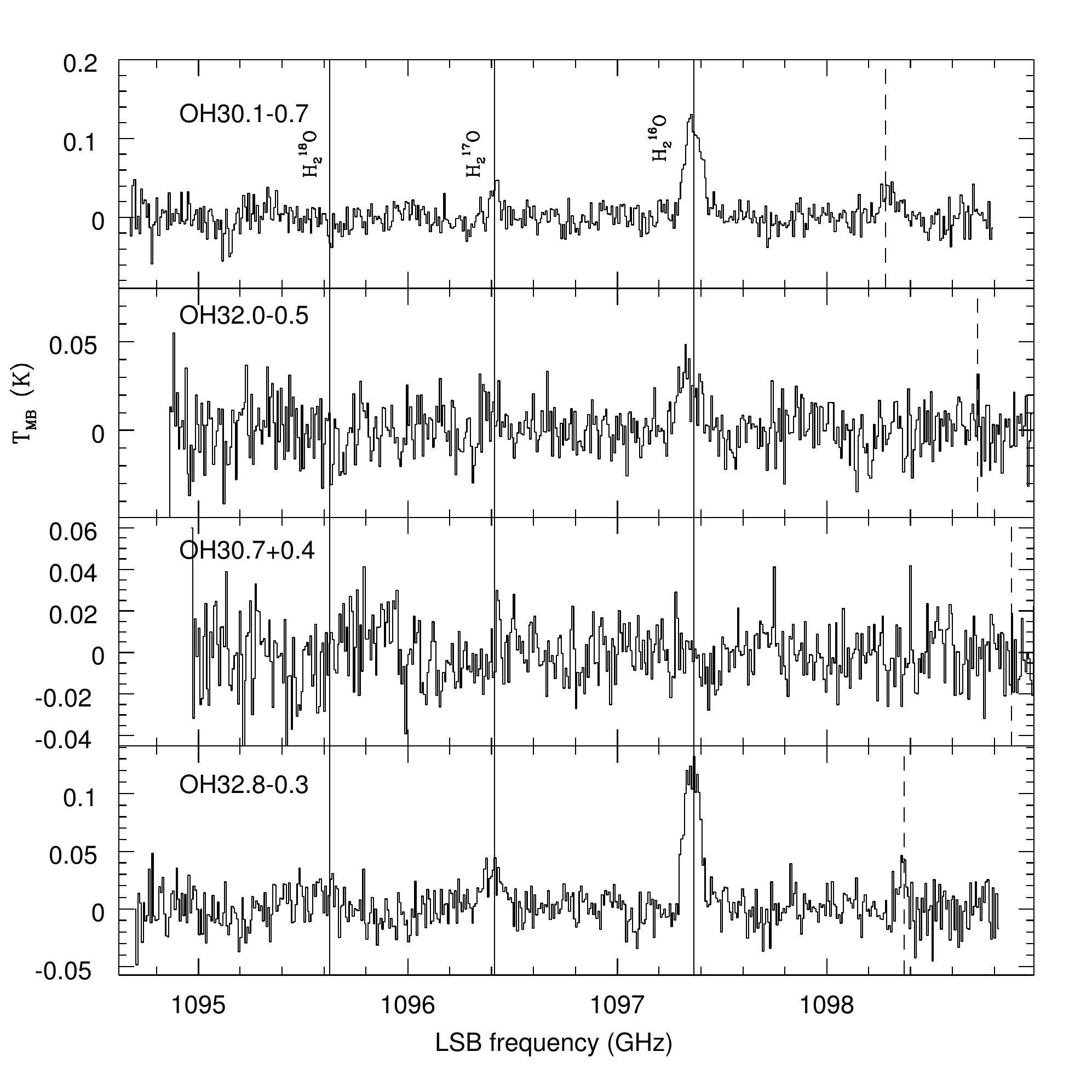}}
\caption{{\it Herschel}-HIFI observation of four extreme OH/IR stars. 
The spectra have 
been corrected for the LSR velocity for each object and rebinned to
$\sim$ 2 km s$^{-1}$ sampling. The vertical lines
indicate the expected position of the H$_{2}^{18}$O, H$_{2}^{17}$O, and 
H$_{2}^{16}$O 3$_{12}$-3$_{03}$ transitions, from left to right. The expected 
position of the H$_{2}^{17}$O 1$_{11}$-0$_{00}$
transition (dashed line) from the upper side-band can 
be seen at the far right.
}
\label{fig_ddt}
\end{figure}

\begin{table*}
\caption{Line fluxes of isotopologues of H$_{2}$O in the observed HIFI
range.}
\label{tab_obs_hifi}
\begin{tabular}{cccrrrrr}
\hline\hline
${\lambda (\mu}$m) & transition & isotope & 
\multicolumn{5}{c}{line fluxes ($\times 10^{-17}$ W m$^{-2}$)} \\
\cline{4-8}
& & & GL5379 & OH26.5 & OH30.1 & OH32.0 & OH32.8 \\
\hline
269.272 & $1_{11}-0_{00}$ & 16 & 18.2 & 8.3  & a    & a    & a    \\
270.774 & $1_{11}-0_{00}$ & 17 &  9.2 & 3.6  & 1.1  & 0.6  & 1.0 \\
272.118 & $1_{11}-0_{00}$ & 18 & <0.8 & <0.6 & a    & a    & a    \\
273.193 & $3_{12}-3_{03}$ & 16 & 29.7 & 11.0 & 4.9  & 2.0  & 4.7 \\
273.430 & $3_{12}-3_{03}$ & 17 &  9.1 & 3.8  & 1.0  & 0.8  & 1.9 \\
273.626 & $3_{12}-3_{03}$ & 18 & <0.8 & <0.6 & <0.4 & <0.3 & <0.4 \\
\hline
\end{tabular}
\tablefoot{a indicates that the transition is not observed. The fluxes for
H$_{2}^{18}$O indicates the upper limit of the detection and as such
can be regarded as an estimated uncertainty of the calculated line fluxes.}
\end{table*}

We obtained SPIRE spectra of eight extreme OH/IR stars and included OH~26.5+0.6
from the archive. The resolution of these spectra is $\sim$ 1.4~GHz,
corresponding to 380 km s$^{-1}$ at a frequency of 1097~GHz
hence the lines are not resolved and 
each observed emission line can be a blend of different molecules.
With the help of the HIFI spectra of OH/IR stars \citep{justtanont12},
we can resolve the contribution of strong emission lines mainly due to H$_{2}$O 
and CO. 

The SPIRE data were reduced using the calibration files SPIRE\_CAL\_12\_2. 
The spectra suffer from high backgrounds due to the location of these
objects in the galactic plane and most stars are thought to have a typical
distance larger than 1 kpc. Interstellar emission lines due to 
[\ion{N}{II}] 
at 205.178~$\mu$m and [\ion{C}{I}] at 370.423 and 609.150~$\mu$m
can be seen in the spectra, with the exception of AFGL~5379 which is thought to
be closer than the other objects. We performed background subtraction from our
spectra by investigating individual off-centre beams and selected those that
have similar background levels as the central beam. However, the region 
between 300-400 $\mu$m is badly affected so that it is difficult to recover
background subtracted data.

\begin{table*}
\caption{Line fluxes of isotopologues of H$_{2}$O in the observed SPIRE
range.}
\label{tab_obs_spire}
\begin{tabular}{lccccccccccc}
\hline\hline
$\lambda (\mu$m) & transition & isotope &
\multicolumn{9}{c}{ Line fluxes ($\times 10^{-17}$ W m$^{-2}$)} \\
\cline{4-12}
& & &                       OH127& GL5379& OH21.5& OH26.5& OH30.7& OH30.1& OH32.0& OH32.8& OH42.3\\
\hline
208.0763&   $7_{26}-6_{33}$& 16&  1.0& 14.1& <0.3\tablefootmark{a}&  5.2&  <0.7\tablefootmark{a}&  1.3&  <0.3\tablefootmark{a}&  0.9&  0.7\\
208.2052&   $7_{26}-6_{33}$& 17&  <0.2\tablefootmark{a}&  <0.8\tablefootmark{a}& <0.3\tablefootmark{a}&  <0.4\tablefootmark{a}&  <0.7\tablefootmark{a}&  <0.3\tablefootmark{a}&  <0.3\tablefootmark{a}&  <0.6\tablefootmark{a}&  <0.4\tablefootmark{a}\\
212.5256&   $5_{23}-5_{14}$& 16&  4.1& 57.9& 1.8& 14.3&  <0.7\tablefootmark{a}&  9.0&  2.0&  5.3&  2.3\\
213.1557&   $5_{23}-5_{14}$& 17&  1.5& 20.2& 1.0&  7.3&  <0.7\tablefootmark{a}&  2.6&  0.8&  2.2&  1.7\\
226.7608&   $6_{25}-5_{32}$& 16&  1.7& 18.9& 0.7&  7.4&  <0.7\tablefootmark{a}&  2.3&  0.5&  2.0&  1.3\\
225.0480&   $6_{25}-5_{32}$& 17&  <0.2\tablefootmark{a}&  <0.8\tablefootmark{a}& <0.3\tablefootmark{a}&  <0.4\tablefootmark{a}&  <0.7\tablefootmark{a}&  <0.3\tablefootmark{a}&  <0.3\tablefootmark{a}&  <0.6\tablefootmark{a}&  <0.4\tablefootmark{a}\\
231.2480\tablefootmark{b}&   $8_{27}-7_{34}$& 16&  1.3& 12.4& <0.3\tablefootmark{a}&  3.1&  <0.7\tablefootmark{a}&  <0.3\tablefootmark{a}&  0.5&  1.3&  2.1\\
233.7153&   $8_{27}-7_{34}$& 17&  <0.2\tablefootmark{a}&  <0.8\tablefootmark{a}& <0.3\tablefootmark{a}&  <0.4\tablefootmark{a}&  <0.7\tablefootmark{a}&  <0.3\tablefootmark{a}&  <0.3\tablefootmark{a}&  <0.6\tablefootmark{a}&  <0.4\tablefootmark{a}\\
234.5306&   $7_{43}-6_{52}$& 16&  0.8&  9.5& <0.3\tablefootmark{a}&  5.0&  <0.7\tablefootmark{a}&  1.2&  0.5&  1.3&  <0.4\tablefootmark{a}\\
226.1507&   $7_{43}-6_{52}$& 17&  <0.2\tablefootmark{a}&  <0.8\tablefootmark{a}& <0.3\tablefootmark{a}&  <0.4\tablefootmark{a}&  <0.7\tablefootmark{a}&  <0.3\tablefootmark{a}&  <0.3\tablefootmark{a}&  <0.6\tablefootmark{a}&  <0.4\tablefootmark{a}\\
243.9740&   $2_{20}-2_{11}$& 16&  2.8& 31.5& 0.7& 11.1&  <0.7\tablefootmark{a}&  4.6&  0.5&  3.4&  <0.4\tablefootmark{a}\\
247.1536&   $2_{20}-2_{11}$& 17&  1.6&  8.7& <0.3\tablefootmark{a}&  3.0&  <0.7\tablefootmark{a}&  0.5&  <0.3\tablefootmark{a}&  <0.6\tablefootmark{a}&  <0.4\tablefootmark{a}\\
248.2468&   $4_{22}-4_{13}$& 16&  3.2& 51.8& 1.4& 10.8&  <0.7\tablefootmark{a}&  6.1&  1.1&  4.3&  1.3\\
250.3256\tablefootmark{b}&   $4_{22}-4_{13}$& 17&  1.3&  5.2& <0.3\tablefootmark{a}&  2.2&  <0.7\tablefootmark{a}&  1.4&  <0.3\tablefootmark{a}&  0.6&  <0.4\tablefootmark{a}\\
257.7948&   $3_{21}-3_{12}$& 16&  4.6& 66.6& 2.4& 16.6&  0.6&  8.7&  2.2&  6.6&  1.2\\
260.9215&   $3_{21}-3_{12}$& 17&  1.1& 11.4& <0.3\tablefootmark{a}&  3.5&  <0.7\tablefootmark{a}&  1.2&  1.4&  0.7&  <0.4\tablefootmark{a}\\
258.8158&   $6_{34}-5_{41}$& 16&  3.3& 11.1& <0.3\tablefootmark{a}&  4.1&  <0.7\tablefootmark{a}&  1.2&  <0.3\tablefootmark{a}&  <0.6\tablefootmark{a}&  <0.4\tablefootmark{a}\\
252.0495&   $6_{34}-5_{41}$& 17&  <0.2\tablefootmark{a}&  <0.8\tablefootmark{a}& <0.3\tablefootmark{a}&  <0.4\tablefootmark{a}&  <0.7\tablefootmark{a}&  <0.3\tablefootmark{a}&  <0.3\tablefootmark{a}&  <0.6\tablefootmark{a}&  <0.4\tablefootmark{a}\\
259.9823&   $3_{12}-2_{21}$& 16&  3.1& 26.8& 0.6&  7.6&  1.7&  3.5&  1.9&  4.1&  1.1\\
256.6420&   $3_{12}-2_{21}$& 17&  2.2& 17.4& <0.3\tablefootmark{a}&  6.2&  <0.7\tablefootmark{a}&  <0.3\tablefootmark{a}&  <0.3\tablefootmark{a}&  1.8&  0.6\\
269.2724&   $1_{11}-0_{00}$& 16&  2.2& 23.2& <0.3\tablefootmark{a}&  8.6&  <0.7\tablefootmark{a}&  3.6&  <0.3\tablefootmark{a}&  1.7&  <0.4\tablefootmark{a}\\
270.7745&   $1_{11}-0_{00}$& 17&  1.4& 11.4& <0.3\tablefootmark{a}&  4.1&  <0.7\tablefootmark{a}&  1.2& <0.3\tablefootmark{a}&  0.8& <0.4\tablefootmark{a}\\
273.1931&   $3_{12}-3_{03}$& 16&  3.6& 36.0& 2.2& 12.1&  <0.7\tablefootmark{a}&  4.1&  1.3&  5.2&  0.9\\
273.4300&   $3_{12}-3_{03}$& 17&  0.5& 12.7& <0.3\tablefootmark{a}&  1.9&  <0.7\tablefootmark{a}&  2.9&  0.6&  <0.6\tablefootmark{a}&  <0.4\tablefootmark{a}\\
303.4562\tablefootmark{b}&   $2_{02}-1_{11}$& 16&  2.5& 22.6& <0.3\tablefootmark{a}&  7.8&  <0.7\tablefootmark{a}&  3.1&  1.0&  2.3&  <0.4\tablefootmark{a}\\
302.3565\tablefootmark{b}&   $2_{02}-1_{11}$& 17&  1.2& 14.9& <0.3\tablefootmark{a}&  5.2&  <0.7\tablefootmark{a}&  3.1&  <0.3\tablefootmark{a}&  2.3& <0.4\tablefootmark{a}\\
308.9641&   $5_{24}-4_{31}$& 16&  0.6&  9.9& <0.3\tablefootmark{a}&  5.6&  <0.7\tablefootmark{a}&  3.0&  <0.3\tablefootmark{a}&  2.8&  2.9\\
303.4711\tablefootmark{b}&   $5_{24}-4_{31}$& 17&  <0.2\tablefootmark{a}&  <0.8\tablefootmark{a}& <0.3\tablefootmark{a}&  <0.4\tablefootmark{a}&  <0.7\tablefootmark{a}&  <0.3\tablefootmark{a}&  <0.3\tablefootmark{a}&  <0.6\tablefootmark{a}&  <0.4\tablefootmark{a}\\
327.2231&   $4_{22}-3_{31}$& 16&  0.8& 13.2& 2.7&  5.9&  <0.7\tablefootmark{a}&  3.7&  4.3&  5.6&  6.0\\
317.2905&   $4_{22}-3_{31}$& 17&  <0.2\tablefootmark{a}&  <0.8\tablefootmark{a}& <0.3\tablefootmark{a}&  <0.4\tablefootmark{a}&  <0.7\tablefootmark{a}&  <0.3\tablefootmark{a}&  <0.3\tablefootmark{a}&  <0.6\tablefootmark{a}&  <0.4\tablefootmark{a}\\
398.6427&   $2_{11}-2_{02}$& 16&  1.4& 20.5& 1.8&  5.9&  1.4&  2.8&  1.9&  2.4&  <0.4\tablefootmark{a}\\
400.5467&   $2_{11}-2_{02}$& 17&  0.7&  6.7& 1.0&  2.3&  1.4&  1.2&  1.4&  1.2&  <0.4\tablefootmark{a}\\
482.9902\tablefootmark{b}&   $5_{32}-4_{41}$& 16&  0.4&  3.2& 0.6&  1.3&  6.7&  <0.3\tablefootmark{a}&  2.1&  0.8&  <0.4\tablefootmark{a}\\
455.2615\tablefootmark{b}&   $5_{32}-4_{41}$& 17&  <0.2\tablefootmark{a}&  <0.8\tablefootmark{a}& <0.3\tablefootmark{a}&  <0.4\tablefootmark{a}&  <0.7\tablefootmark{a}&  <0.3\tablefootmark{a}&  <0.3\tablefootmark{a}&  <0.6\tablefootmark{a}&  <0.4\tablefootmark{a}\\
538.2890&   $1_{10}-1_{01}$& 16&  0.9&  6.9& <0.3\tablefootmark{a}&  1.3&  <0.7\tablefootmark{a}&  0.8&  0.7&  0.6&  <0.4\tablefootmark{a}\\
543.0817\tablefootmark{b}&   $1_{10}-1_{01}$& 17&  0.8&  4.6& <0.3\tablefootmark{a}&  <0.4\tablefootmark{a}&  0.7&  0.5&  0.7&  <0.6\tablefootmark{a}&  <0.4\tablefootmark{a}\\
\hline
\end{tabular}
\tablefoot{
\tablefoottext{a}{an upper limit of the detection.}
\tablefoottext{b}{possible blend with another line.}
}
\end{table*} 

A number of the stars observed with SPIRE have been observed
using the PACS instrument. Together with the PACS data taken by the 
MESS guaranteed time program \citep{groen11} for AFGL~5379
and from another open-time program (PI. M.J. Barlow) for OH~26.5+0.6,
we have full spectral 
coverage from 50 to 670~$\mu$m for all the stars in our sample.
For OH~127.8+0.0, the data were taken as part of the calibration time
\citep{lombaert13}.
We used the calibration files PACS\_CAL\_48\_0 for our targets. The flux from
the central 
3$\times$3 spaxels were extracted. We note that the data suffer from significant 
leakage in the red part of the spectrum, which means that the data between 
95-100~$\mu$m and beyond
$\sim$ 190~$\mu$m cannot be recovered. 

In the PACS spectral range, we identify the emission lines as
coming from H$_{2}$O, and CO plus three sets of lines due to OH at 79, 119, and
163~$\mu$m (Fig.~\ref{gl5379_pacs} and Appendix~\ref{app_linefluxes}). 
These lines 
have previously been reported by \cite{sylvester97} and \cite{lombaert13} 
and are thought to be
the pumping line for the OH masers seen in these objects. Although 
the archived spectra from the short-wavelength spectrometer  
\citep[LWS,][]{clegg96} aboard the Infrared
Space Observatory \citep[ISO,][]{kessler96} of some of these stars
are very noisy, there may be a possible hint of an absorption of the infrared
pumping line at 53~$\mu$m. The other infrared pumping line at 34.6~$\mu$m
was first reported towards two supergiants \citep{justtanont96,
sylvester97},
but has not been detected towards AGB stars observed with ISO.
No strong emission lines due to other molecules apart from 
H$_{2}$O, OH, and CO have 
been reported from OH~127.8+0.0 \citep{lombaert13}.

\begin{table*}
\caption{Line fluxes of isotopologues of H$_{2}$O in the observed PACS
range.}
\label{tab_obs_pacs}
\begin{tabular}{lcccccccc}
\hline\hline
$\lambda (\mu$m) & transition & isotope &
\multicolumn{6}{c}{ Line fluxes ($\times 10^{-17}$ W m$^{-2}$)} \\
\cline{4-9}
& & &                          GL5379& OH26.5& OH30.1& OH32.0& OH32.8& OH42.3\\
\hline
57.6365&  $4_{22}-3_{13}$ &16& 115.2&  57.8& 10.6&  <1.6\tablefootmark{a}&  4.6&  <2.7\tablefootmark{a}\\
57.7447\tablefootmark{b}&  $4_{22}-3_{13}$ &17&  <36.5\tablefootmark{a}&  41.1& <4.4\tablefootmark{a}&  <1.6\tablefootmark{a}& <3.1\tablefootmark{a}&  <2.7\tablefootmark{a}\\
67.0892\tablefootmark{b}&  $3_{31}-2_{20}$ &16& 133.5&  79.1& 16.3&  5.7& 19.0&  4.6\\
67.5081\tablefootmark{b}&  $3_{31}-2_{20}$ &17&   <13.8\tablefootmark{a}&  16.0& <4.4\tablefootmark{a}&  <1.6\tablefootmark{a}&  <3.1\tablefootmark{a}&  <2.0\tablefootmark{a}\\
75.3807&  $3_{21}-2_{12}$ &16& 135.3&  55.7& 13.7&  3.9&  8.2&  7.3\\
75.6381&  $3_{21}-2_{12}$ &17&  <25.9\tablefootmark{a}&  28.3& <4.4\tablefootmark{a}&  3.3&  0.4&  <1.9\tablefootmark{a}\\
75.9099&  $5_{50}-5_{41}$ &16& 177.1&  74.1& 23.7&  8.6& 20.9&  7.7\\
76.7892&  $5_{50}-5_{41}$ &17&  <25.9\tablefootmark{a}&  20.0& <4.4\tablefootmark{a}&  <1.6\tablefootmark{a}&  <2.1\tablefootmark{a}&  <1.9\tablefootmark{a}\\
89.9884&  $3_{22}-2_{11}$ &16& 254.5&  84.0& 33.3& 12.4& 22.5& 12.9\\
90.4885&  $3_{22}-2_{11}$ &17&  <24.7\tablefootmark{a}&  14.2&  7.9&  2.5&  <2.1\tablefootmark{a}&  6.6\\
108.0732&  $2_{21}-1_{10}$ &16& 100.1&  34.0&  9.5&  3.7&  6.6&  2.0\\
108.7449&  $2_{21}-1_{10}$ &17&  19.4&  19.5&  3.1&  2.8&  4.0&  <0.7\tablefootmark{a}\\
121.7217&  $4_{32}-4_{23}$ &16& 227.9&  70.5& 25.4&  7.7& 17.2&  7.0\\
122.8999&  $4_{32}-4_{23}$ &17&  13.6&  10.9&  4.2&  2.1&  2.5&  <0.4\tablefootmark{a}\\
132.4084&  $4_{23}-4_{14}$ &16& 176.8&  50.6& 17.8&  8.3& 13.2&  3.5\\
133.0943&  $4_{23}-4_{14}$ &17&  31.0&  11.6&  3.3&  1.6&  2.3&  <0.4\tablefootmark{a}\\
134.9353&  $5_{14}-5_{05}$ &16& 109.2&  36.7& 13.2&  5.4&  9.4&  3.4\\
134.7376&  $5_{14}-5_{05}$ &17&  15.5&   9.9& <1.0\tablefootmark{a}&  <0.5\tablefootmark{a}&  2.3&  <0.4\tablefootmark{a}\\
136.4960&  $3_{30}-3_{21}$ &16& 133.9&  37.1& 12.5&  5.8&  9.1&  3.0\\
138.2514&  $3_{30}-3_{21}$ &17&  22.4&  12.9&  1.5&  <0.5\tablefootmark{a}&  1.7&  1.7\\
138.5278&  $3_{13}-2_{02}$ &16& 128.1&  45.0& 13.7&  6.0&  9.7&  3.5\\
139.0864&  $3_{13}-2_{02}$ &17&  28.9&  17.2&  3.8&  2.8&  3.0&  <0.4\tablefootmark{a}\\
156.1940\tablefootmark{b}&  $3_{22}-3_{13}$ &16& 184.1&  63.7& 20.9& 10.0& 13.8&  5.0\\
157.2836&  $3_{22}-3_{13}$ &17&  22.0&  11.9&  1.8&  <0.9\tablefootmark{a}&  <0.9\tablefootmark{a}&  <0.5\tablefootmark{a}\\
156.2652\tablefootmark{b}&  $5_{23}-4_{32}$ &16&  21.1&  21.5&  2.1&  3.5&  2.1&  1.0\\
153.8758&  $5_{23}-4_{32}$ &17& <5.6\tablefootmark{a}& <4.1\tablefootmark{a}&  <1.0\tablefootmark{a}&  <0.9\tablefootmark{a}&  <0.6\tablefootmark{a}&  <0.5\tablefootmark{a}\\
179.5267&  $2_{12}-1_{01}$ &16& 122.8&  46.8& 10.7&  5.9&  2.4&  <0.5\tablefootmark{a}\\
180.3302&  $2_{12}-1_{01}$ &17&  45.7&  32.3&  5.3&  <0.9\tablefootmark{a}&  0.7&  <0.5\tablefootmark{a}\\
180.4883&  $2_{21}-2_{12}$ &16& 112.3&  39.0& 11.2&  4.4&  1.3&  2.0\\
182.0899&  $2_{21}-2_{12}$ &17&  41.5&  15.2&  <1.0\tablefootmark{a}&  <0.9\tablefootmark{a}&  <0.6\tablefootmark{a}&  1.7\\
\hline
\end{tabular}
\tablefoot{
\tablefoottext{a}{an upper limit of the detection.}
\tablefoottext{b}{a possible blend with higher excited H$_{2}$O line.}
}
\end{table*}

\section{Discussion}

\cite{justtanont13} found strong H$_{2}^{16}$O and H$_{2}^{17}$O emission
lines in two extreme OH/IR stars, AFGL~5379 and OH~26.5+0.6, while there is 
no detectable emission due to H$_{2}^{18}$O. 
This is
in contrast to what is observed in the interstellar medium and the Sun
\citep{wilson94} where $^{18}$O is more abundant than $^{17}$O. 
The question arises
if other extreme OH/IR stars show the same behaviour.

\subsection{Resolved HIFI spectra}

In four objects, we obtained HIFI spectra that cover the frequency range of
the 3$_{12}$-3$_{03}$ line for all three isotopologues. From Fig.~\ref{fig_ddt},
we detect the main line at 1097.365~GHz (273.200~$\mu$m) in all the sources
except OH~30.7+0.4.
The expansion velocity of these stars is typically 15 km s$^{-1}$, hence
with an observed velocity resolution of 1 km s$^{1}$
the line is well resolved.
The spectrum of OH~32.0-0.5 is too noisy to
confirm the detection of H$_{2}^{17}$O at 1096.414~GHz (273.430~$\mu$m);
H$_{2}^{17}$O can be seen in OH~30.1-0.7 and OH~32.8-0.3, along with
the H$_{2}^{17}$O 1$_{11}$-0$_{00}$ line from the upper side-band.
No evidence of the H$_{2}^{18}$O 3$_{12}$-3$_{03}$ line at 1095.627~GHz
(273.626~$\mu$m) is seen in our spectra (Table~\ref{tab_obs_hifi}). 
This is similar to the
non-detection of H$_{2}^{18}$O in AFGL~5379 and OH~26.5+0.6
where both H$_{2}^{16}$O and H$_{2}^{17}$O lines are clearly detected
\citep{justtanont13}.

Based on these observations, we searched for the presence of all isotopologues 
of H$_{2}$O in the SPIRE and PACS spectra. 

\subsection{SPIRE spectra}

In these spectra, we are able to
discern a number of possible H$_{2}^{17}$O lines. The SPIRE spectrum 
of AFGL~5379 (Figs.~\ref{gl5379_spire_sw} and \ref{gl5379_spire_lw}) shows
the identification of both isotopologues, along with other molecules
such as CO, SiO, HCN, and H$_{2}$S (see Appendix~\ref{app_spire})
although the lines are not resolved owing
to the poor spectral resolution of the SPIRE instrument
($\lambda/\Delta\lambda \sim$ 370-1290 for 670-194~$\mu$m). In order to
calculate the line fluxes, we employed the special script in the HIPE 
data reduction package written
for unapodized SPIRE data (SPIRE\_linefitting.py), 
taking into account the fit to the sinc function
of the line profiles. For display purposes, we show the
apodized spectra (with the sinc function corrected) together with
the calculated Gaussian line profiles because the lines 
are unresolved, with a width of 0.078 cm$^{-1}$ for unapodized spectra,
i.e. $F_{\rm line} = 1.08 \times F_{\rm peak} \times$ FWHM.
However, one caveat of the derived line fluxes
is that the decomposed molecular components depend only on the central 
frequencies of the lines and not on the expected transition line strengths.
For this purpose we include species thath exhibit several transitions and 
select lines within the SPIRE range with an upper energy $\leq$ 1000~K.
We did, however, add an exception for the H$_{2}$O $\nu_{2}$ transitions as
these lines can sometimes be bright when they are masing.
We listed the pair of detected H$_{2}^{16}$O and H$_{2}^{17}$O fluxes in
Table~\ref{tab_obs_spire}. The estimated uncertainty for unblended lines
is 30\%, increasing to $\sim$ 50\% for blended lines.
For our purpose, blended lines are defined as lines that have
two or more species that are separated by less than the spectral
resolution of the instrument and so the peaks are indistinguishable.
In many cases, there are lines that overlap with distinct peaks
which makes the line flux determination cleaner than the blended lines.
We note that a series of lines due to H$_{2}$S are detected in the
SPIRE wavelength range for most of our objects. 
The full list of all the lines detected and plots of the SPIRE spectra
observed can be found in Appendix~\ref{app_spire}  and 
\ref{app_spire_plots}, respectively.

\subsection{PACS spectra}

For the PACS spectra, we fitted a Gaussian to individual lines to derive
a line flux as the lines are also unresolved with a resolving power of
1000-5000, corresponding to a velocity resolution of
$\sim$ 300-60 km s$^{-1}$,
for the long and short wavelengths, respectively. 
Since most of the lines are due to H$_{2}$O, we decided to fit 
these before attributing the unfitted lines to other molecules such as CO and 
H$_{2}$S. Figure~\ref{gl5379_pacs} shows the continuum subtracted spectrum of
AFGL~5379. Unfortunately, the source was not at the central spaxel when it
was observed. This resulted in a loss of flux. However, we
corrected this based on the archived ISO-LWS continuum flux data. 
The PACS data
have been multiplied by a factor of 1.78 to get the flux to agree with
the ISO-LWS flux. Another artefact is that the shortest wavelength part of the 
PACS spectrum cannot be recovered as it is badly affected by the 
foreground and background interstellar [\ion{O}{I}]~63 emission.
We did not detect any emission lines in the PACS spectrum of OH~21.5+0.5.
The PACS spectrum of OH~30.7+0.4 does not show definitive detections of
H$_{2}^{17}$O lines. These two objects are hence not listed in 
Table~\ref{tab_obs_pacs}.

The line fluxes of H$_{2}^{16}$O and H$_{2}^{17}$O are 
listed in Table~\ref{tab_obs_pacs}. The estimated uncertainty for each derived
line flux depends on the errors in baseline subtraction and the
rms noise of the spectrum which can affect the flux by $\sim$ 30\%.
Two H$_{2}^{17}$O lines at 57.74 and 67.51~$\mu$m are close in wavelength 
to much higher
excited lines of H$_{2}$O 8$_{17}$-7$_{26}$ and 11$_{48}$-11$_{39}$,
with upper energy levels of 1270K and 2652K, respectively),
so we do not expect the calculated line fluxes
to be much affected. We note here that although Tables~\ref{tab_obs_spire}
and \ref{tab_obs_pacs} 
list only pairs of detected 
H$_{2}^{16}$O and H$_{2}^{17}$O, many H$_{2}^{16}$O lines with no
accompanying less abundant isotopologue are detected.
The full list of all the lines detected and plots of the PACS spectra
observed can be found in Appendix~\ref{app_linefluxes}
and \ref{app_pacs_plots}, respectively.

\begin{figure*}
\centering
\includegraphics[width=17cm]{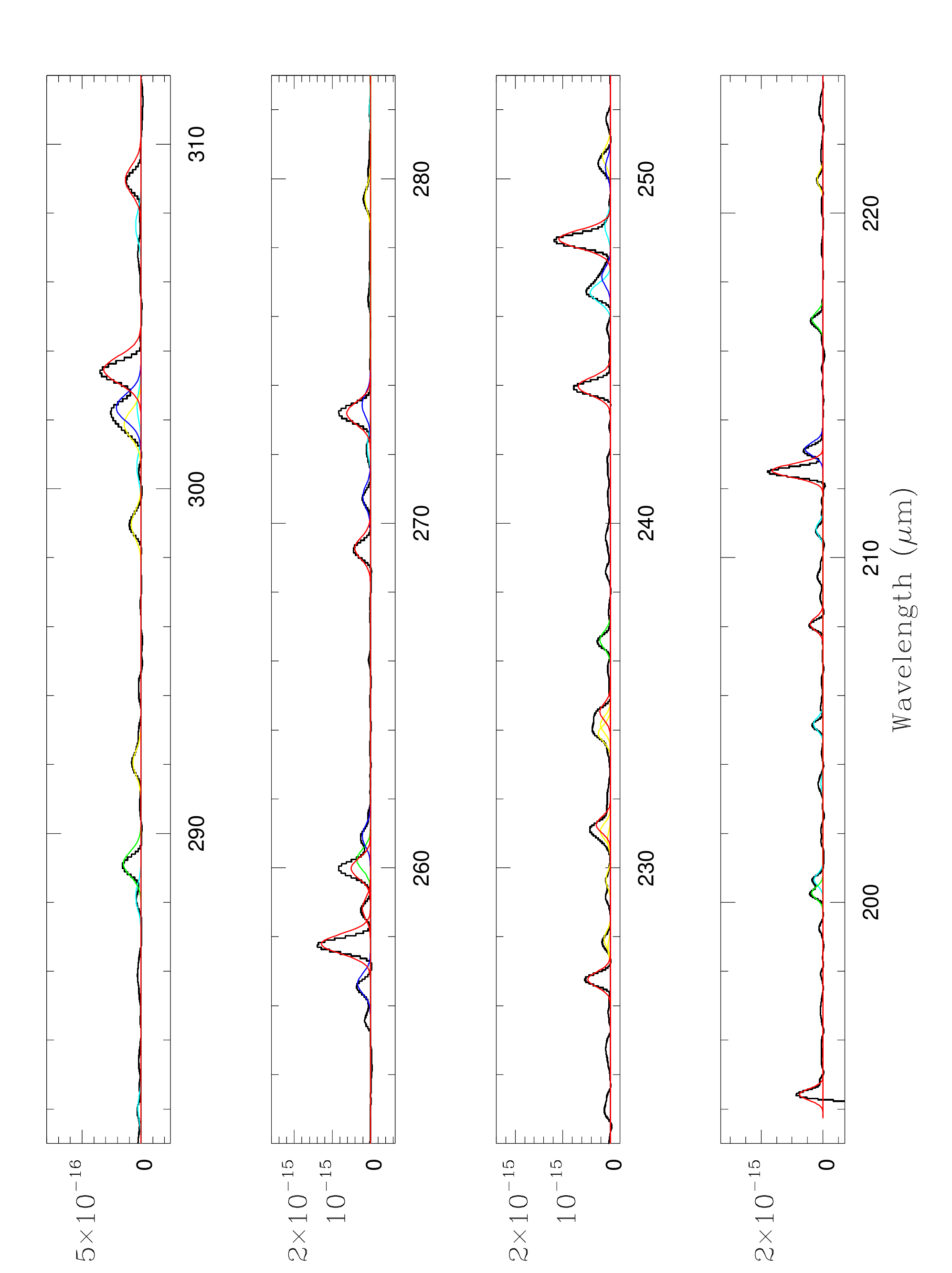}
\caption{The continuum subtracted apodized SPIRE spectrum of
AFGL~5379 (histogram)
with the Gaussian fits for H$_{2}$O (red),  H$_{2}^{17}$O (blue),
CO (green), and H$_{2}$S (yellow). Other molecules such as SiO, HCN, and the
interstellar lines of [\ion{C}{I}] and [\ion{N}{II}] are shown in cyan.
The flux for SPIRE and PACS spectra is in W m$^{-2} \mu$m$^{-1}$.}
\label{gl5379_spire_sw}
\end{figure*}

\begin{figure*}
\centering
\includegraphics[width=17cm]{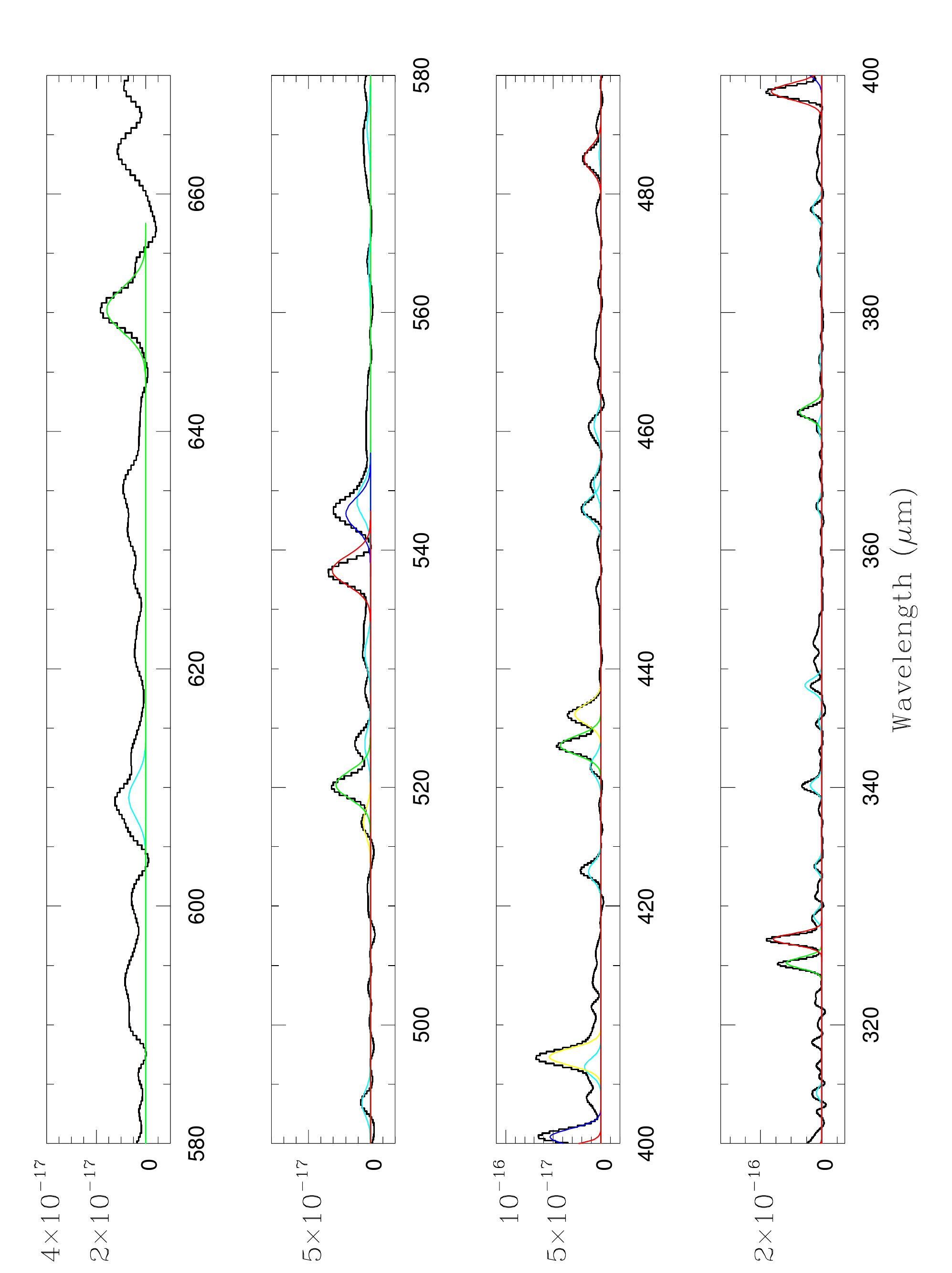}
\caption{The continuum subtracted apodized  SPIRE spectrum of 
AFGL~5379 (histogram)
with the Gaussian fits for H$_{2}$O (red),  H$_{2}^{17}$O (blue),
CO (green), and H$_{2}$S (yellow). Other molecules are shown in cyan.}
\label{gl5379_spire_lw}
\end{figure*}

\begin{figure*}
\centering
\includegraphics[width=17cm]{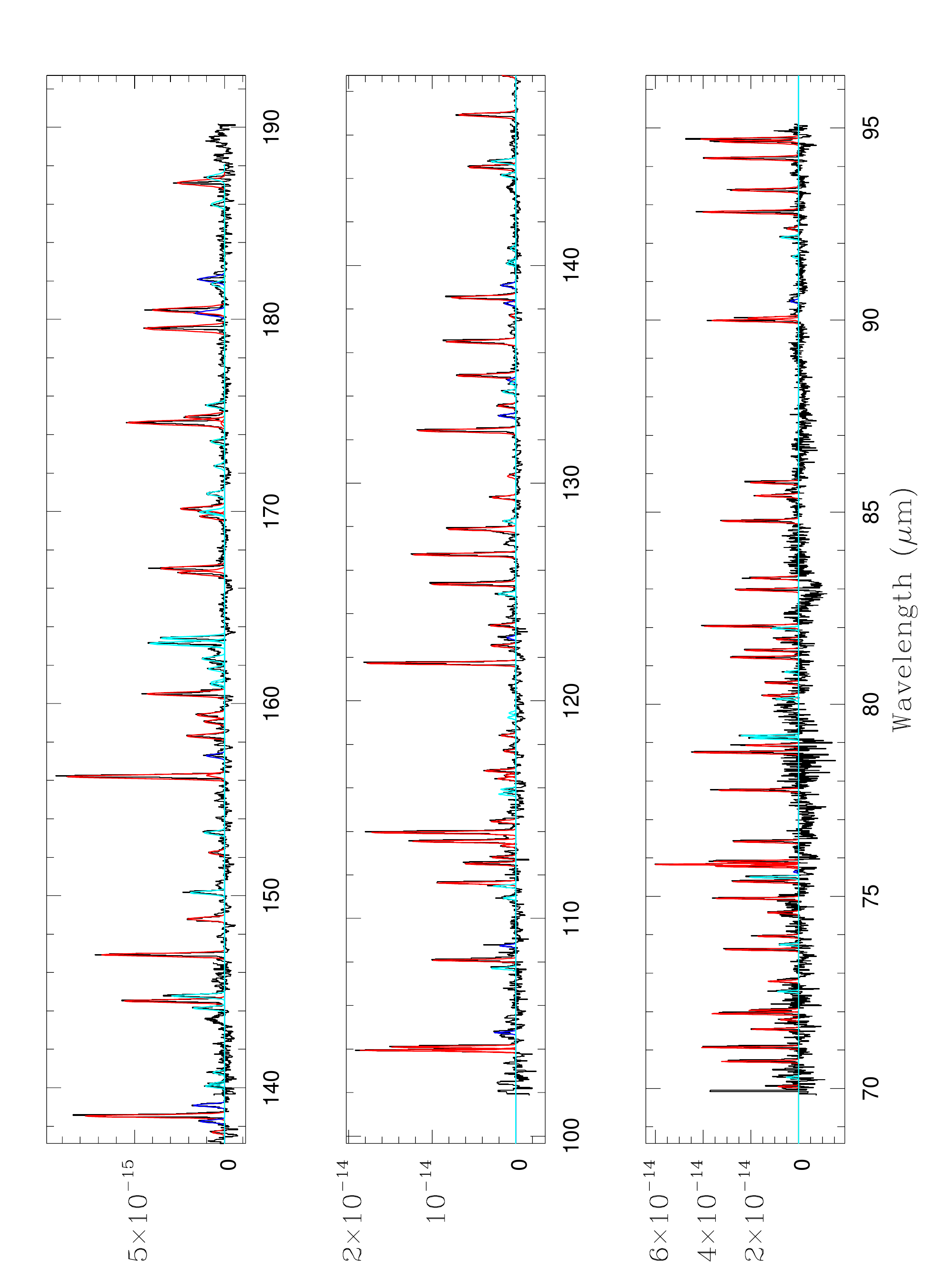}
\caption{The continuum subtracted PACS spectrum of AFGL~5379 (histogram)
corrected to the ISO-LWS flux level
with the Gaussian fits for H$_{2}$O (red) and H$_{2}^{17}$O (blue).
Other molecules are shown in cyan.} 
\label{gl5379_pacs}
\end{figure*}

\subsection{H$_{2}$O isotopologues}

\begin{figure}[t]
\centering
\resizebox{\hsize}{!}{\includegraphics{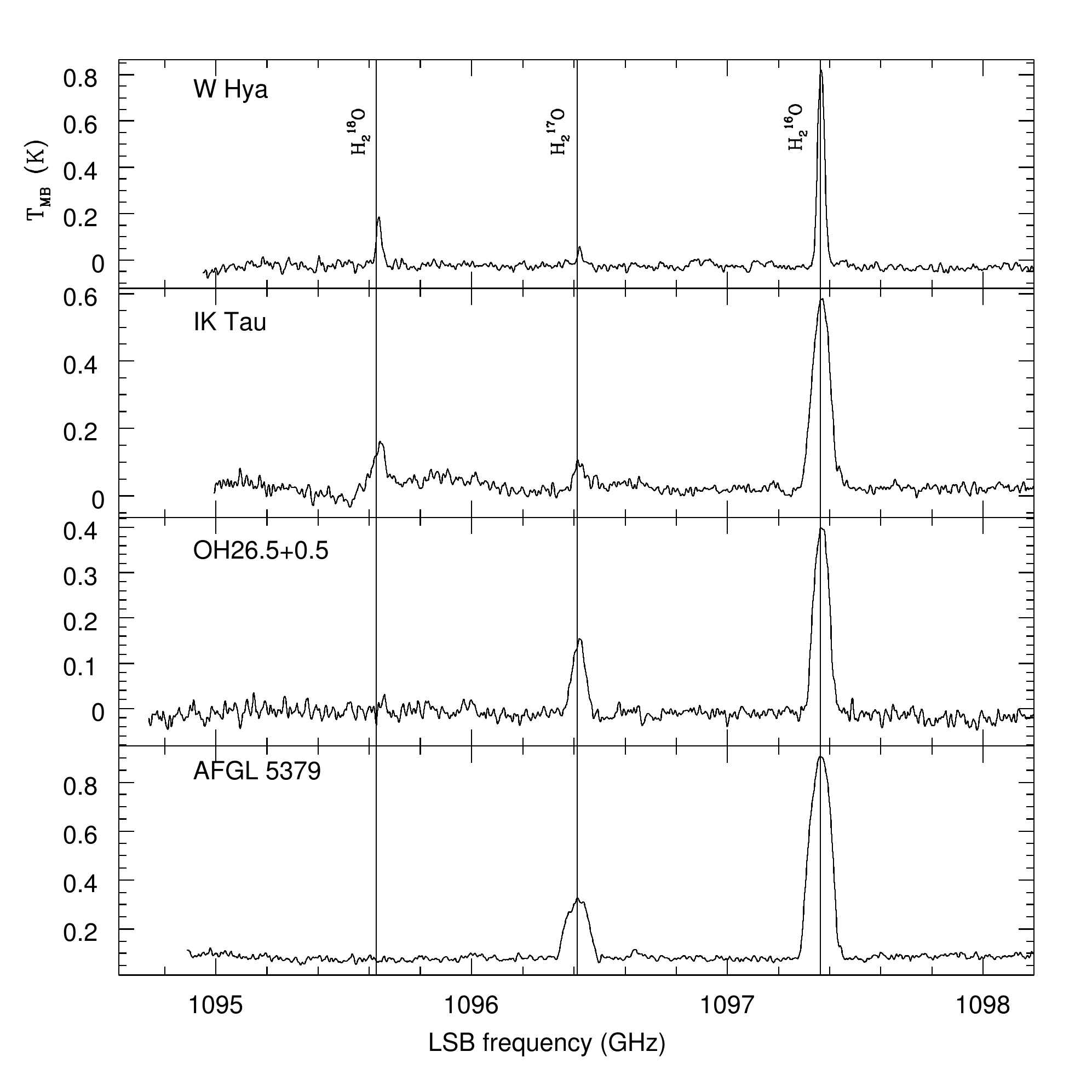}}
\caption{Observations taken from the HIFISTARS sample
showing the 3$_{12}$-3$_{03}$ transition of ortho-H$_{2}$O
(see Fig~\ref{fig_ddt}).
The H$_{2}^{18}$O line is clearly detected
in W Hya and IK Tau \citep{justtanont12} which have much lower mass-loss rates
than in AFGL~5379 and OH~26.5+0.6 \citep{justtanont13}.
}
\label{fig_hifistars}
\end{figure}

In all cases, we also searched for the presence of H$_{2}^{18}$O in the HIFI, 
SPIRE and PACS spectra. 
It is clear that in the HIFI spectra where
line blending is not an issue,  the $3_{12}-3_{03}$ transition at 1095.627~GHz
is below the noise limit (Fig.~\ref{fig_ddt}). In PACS and SPIRE spectra,
portions containing isolated unblended transition with another possible 
lines have been carefully looked at, but there is no emission detected
above the noise. With these results, we conclude that our sample of
extreme OH/IR star spectra lacks the presence of H$_{2}^{18}$O. 
The upper limit of the H$_{2}^{18}$O/H$_{2}^{17}$O line ratios  
are given in Appendix~\ref{app_flux_ratios}. Here, it is clear that the ratios are below 
unity in cases where H$_{2}^{17}$O lines are detected above the noise. 
In the sample of O-rich AGB stars from the guaranteed time program HIFISTARS 
with a lower mass-loss rate,
with the same frequency settings as for the OH/IR stars observed in our
open-time program \citep{justtanont12}, the line fluxes of H$_{2}^{18}$O
are always brighter than those of H$_{2}^{17}$O for both the 
$1_{11}-0_{00}$ and $3_{12}-3_{03}$ transitions (Fig.~\ref{fig_hifistars}).

The absence of H$_{2}^{18}$O in these stars is counter-intuitive
considering the observed isotopic ratio of $^{18}$O/$^{17}$O $\sim$ 3 in
the interstellar medium. However, calculations of nucleosynthesis
during the AGB phase for intermediate-mass stars predict that for stars with 
an initial mass larger than 5~M$_{\odot}$, the temperature at the base
of the convective layer is high enough to start hot-bottom burning,
preventing the star from becoming a
carbon star \citep[see e.g.,][]{lattanzio96,lattanzio03}. 
At the beginning of hot-bottom burning, $^{18}$O is destroyed while the
production of 
$^{17}$O is increased by an order of magnitude,
hence the $^{18}$O/$^{17}$O ratio has an expected value of 
10$^{-6}$ while the $^{16}$O/$^{17}$O ratio is expected to be 
$\leq$ 350 \citep{lattanzio96}.
Hot-bottom burning will finally cease when the star loses most of its mass
such that the envelope mass is below 1~M$_{\odot}$. However,
the third dredge-up can still continue and will change the 
$^{12}$C/$^{13}$C ratio while leaving isotopic ratios of other
elements almost unaffected. Studies of pre-solar grains
reveal very few grains with extremely low $^{18}$O/$^{16}$O ratios,
which can possibly come from intermediate-mass AGB stars \citep{lugaro07,
nittler10b}, while most of the grains show oxygen isotopic ratios
commonly expected from low-mass stars. The rarity of these
presolar grains with low $^{18}$O content is consistent with the
expected population of intermediate-mass stars assuming an
initial mass function 
\citep{salpeter55, scalo86}
and the relatively short lifetimes of such stars.

The line flux ratios of H$_{2}^{16}$O/H$_{2}^{17}$O vary between about
unity and less than 10 where the corresponding transition of both
molecules is detected (Tables~\ref{tab_obs_spire} and \ref{tab_obs_pacs}). 
Most of our reliable line flux ratios from HIFI give values between 2 and 5.
This clearly indicates that at least the main
line is optically thick. 
In order to derive isotopic abundance ratio
from our observations,
a radiative transfer calculation must be performed, which will be
addressed in a forthcoming paper.

\section{Summary}

The H$_{2}$O line fluxes observed with {\it Herschel} are presented for
a sample of nine extreme OH/IR stars. These stars are close to the 
galactic plane and are thought to be population I stars. They all
show strong H$_{2}$O emission from the main isotopologue and
from H$_{2}^{17}$O. The absence of H$_{2}^{18}$O detection was unexpected 
considering the solar and galactic ratio of $^{18}$O/$^{17}$O of 3-5
\citep{wilson94,wouterloot08}.

To explain this question, we propose that our sample stars have
undergone hot-bottom burning, which preferentially destroys $^{18}$O
relative to the other two isotopes. 
During hot-bottom burning, the abundance of 
$^{17}$O are expected to go up by an order of
magnitude while the $^{18}$O abundance drops by more than two orders
of magnitude.
For such a process to happen, the
bottom of the convective layer is required to be hotter than 80$\times
10^{6}$~K. This high temperature can be achieved in stars with initial
masses of at least 5~M$_{\odot}$ \citep{karakas14}.
It should be noted that hot-bottom burning ceases when the star loses
sufficient mass that the high temperature cannot be maintained.
Although the isotopic ratios of most elements remain the same after
this cessation, the $^{12}$C abundance can increase thanks to the continuation
of the third dredge-up process bringing up carbon made by the
triple-alpha reaction.
The materials expelled from these stars will have an impact on  
local isotopic ratios and may also affect the overall
chemical evolution of the Galaxy.

The {\it Herschel} observations of OH/IR stars complement 
previous optical ground based AGB star observations of the 
$^{7}$Li line by \cite{garcia13},
which provides another indication of the operation
of the hot-bottom burning in intermediate-mass stars. 
It may also be possible to search for signatures of hot-bottom
burning using elements synthesized during this phase, such as
$^{22}$Ne and $^{25}$Mg. Observations of isotopic ratios of various
elements together with theoretical calculations of nucleosynthesis
can yield better constraints on the initial mass of these stars.

\begin{acknowledgements}
This research is partly funded by the Swedish National Space Board. 
We also thank both the referee and the editor 
for further comments for improvement of this paper.

HIFI has been designed and built by a consortium of institutes and university 
departments from across Europe, Canada and the United States under the 
leadership of SRON Netherlands Institute for Space Research, Groningen, The 
Netherlands and with major contributions from Germany, France and the US. 
Consortium members are: Canada: CSA, U.Waterloo; France: CESR, LAB, LERMA, 
IRAM; Germany: KOSMA, MPIfR, MPS; Ireland, NUI Maynooth; Italy: ASI, IFSI-INAF,
Osservatorio Astrofisico di Arcetri-INAF; Netherlands: SRON, TUD; Poland: CAMK,
CBK; Spain: Observatorio Astronómico Nacional (IGN), Centro de Astrobiología
(CSIC-INTA). Sweden: Chalmers University of Technology - MC2, RSS \& 
GARD; Onsala Space Observatory; Swedish National Space Board, Stockholm 
University - Stockholm Observatory; Switzerland: ETH Zurich, FHNW; USA: 
Caltech, JPL, NHSC.

PACS has been developed by a consortium of institutes led by MPE (Germany) 
and including UVIE (Austria); KU Leuven, CSL, IMEC (Belgium); CEA, LAM 
(France); MPIA (Germany); INAF-IFSI/OAA/OAP/OAT, LENS, SISSA (Italy); IAC 
(Spain). This development has been supported by the funding agencies BMVIT 
(Austria), ESA-PRODEX (Belgium), CEA/CNES (France), DLR (Germany), ASI/INAF 
(Italy), and CICYT/MCYT (Spain).

SPIRE has been developed by a consortium of institutes led by Cardiff 
University (UK) and including Univ. Lethbridge (Canada); NAOC (China); 
CEA, LAM (France); IFSI, Univ. Padua (Italy); IAC (Spain); Stockholm 
Observatory (Sweden); Imperial College London, RAL, UCL-MSSL, UKATC, 
Univ. Sussex (UK); and Caltech, JPL, NHSC, Univ. Colorado (USA). This 
development has been supported by national funding agencies: CSA 
(Canada); NAOC (China); CEA, CNES, CNRS (France); ASI (Italy); MCINN 
(Spain); SNSB (Sweden); STFC and UKSA (UK); and NASA (USA).

\end{acknowledgements}
\bibliographystyle{aa}
\bibliography{ref}

\appendix
\onecolumn

\section{Line flux ratios}
\label{app_flux_ratios}

This section of the Appendix shows the line flux ratios of 
H$_{2}^{18}$O/H$_{2}^{17}$O. We note that since the H$_{2}^{18}$O is not detected,
we give lower limits to the line ratios. In the case when both
isotopologues are not detected, the ratios are set to 1.0.

\begin{table*}
\caption{Line flux ratios of H$_{2}^{18}$O/H$_{2}^{17}$O observed by
{\it Herschel}-HIFI.}
\label{tab_18rat_hifi}
\begin{tabular}{crrrrr}
\hline\hline
transition  & GL5379 & OH26.5 & OH30.1 & OH32.0 & OH32.8 \\
\hline
$1_{11}-0_{00}$ &  <0.1 & <0.2 & a    & a    & a    \\
$3_{12}-3_{03}$ &  <0.1 & <0.2 & <0.4 & <0.4 & <0.2 \\
\hline
\end{tabular}
\tablefoot{a indicates that the transition is not observed.} 
\end{table*}

\begin{table*}
\caption{Line flux ratios of H$_{2}^{18}$O/H$_{2}^{17}$O in the observed PACS
range.}
\label{tab_18rat_pacs}
\begin{tabular}{crrrrrr}
\hline\hline
 transition &     GL5379& OH26.5& OH30.1& OH32.0& OH32.8& OH42.3\\
\hline
 $ 4_{22}-3_{13}$ &   --   & <0.3  &  --    &  --    &  --    &  --   \\
 $ 3_{31}-2_{20}$ &   --   & <0.8  &  --    &  --    &  --    &  --   \\
 $ 3_{21}-2_{12}$ &   --   & <0.4  &  --    & <0.5   & <0.6   &  --   \\
 $ 5_{50}-5_{41}$ &   --   & <0.6  &  --    &  --    &  --    &  --   \\
 $ 3_{22}-2_{11}$ &  <0.7  & <0.9  &  --    & <0.6   &  --    & <0.3  \\
 $ 2_{21}-1_{10}$ &  <0.3  & <0.4  & <0.3   & <0.2   & <0.2   &  --   \\
 $ 4_{32}-4_{23}$ &  <0.4  & <0.6  & <0.2   & <0.2   & <0.4   &  --   \\
 $ 4_{23}-4_{14}$ &  <0.2  & <0.6  & <0.3   & <0.3   & <0.4   &  --   \\
 $ 5_{14}-5_{05}$ &  <0.4  & <0.7  &  --    &  --    & <0.4   &  --   \\
 $ 3_{30}-3_{21}$ &  <0.3  & <0.5  & <0.7   &  --    & <0.5   & <0.2  \\
 $ 3_{13}-2_{02}$ &  <0.2  & <0.4  & <0.3   & <0.2   & <0.3   &  --   \\
 $ 3_{22}-3_{13}$ &  <0.3  & <0.6  & <0.6   &  --    &  --    &  --   \\
 $ 5_{23}-4_{32}$ &   --   &  --   &  --    &  --    &  --    &  --   \\
 $ 2_{12}-1_{01}$ &  <0.1  & <0.2  & <0.2   &  --    & <0.9   &  --   \\
 $ 2_{21}-2_{12}$ &  <0.1  & <0.5  &  --    &  --    &  --    & <0.3  \\
\hline
\end{tabular}
\tablefoot{The "--" symbol indicates that both H$_{2}^{18}$O and 
H$_{2}^{17}$O are not detected.}
\end{table*}

\begin{table*}
\caption{Ratios of H$_{2}^{18}$O/H$_{2}^{17}$O line fluxes observed with SPIRE.
}
\label{tab_18rat_spire}
\begin{tabular}{crrrrrrrrr}
\hline\hline
transition &     OH127& GL5379& OH21.5& OH26.5& OH30.7& OH30.1& OH32.0& OH32.8& 
OH42.3\\
\hline
$7_{26}-6_{33}$&  --  & --   &  --  &  --  &  --  & --   &  --  &  --  & --\\
$5_{23}-5_{14}$& <0.1 & <0.1 &  --  & <0.1 &  --  & <0.1 & <0.4 & <0.3 &<0.2\\
$6_{25}-5_{32}$&  --  & --   &  --  &  --  &  --  & --   &  --  &  --  & --\\
$8_{27}-7_{34}$&  --  & --   &  --  &  --  &  --  & --   &  --  &  --  & --\\
$7_{43}-6_{52}$&  --  & --   &  --  &  --  &  --  & --   &  --  &  --  & --\\
$2_{20}-2_{11}$& <0.1 & <0.1 &  --  & <0.1 &  --  & <0.6 &  --  &  --  & --\\
$4_{22}-4_{13}$& <0.2 & <0.2 &  --  & <0.2 &  --  & <0.2 &  --  & <1.0 & --\\
$3_{21}-3_{12}$& <0.2 & <0.1 &  --  & <0.1 &  --  & <0.3 & <0.2 & <0.9 & --\\
$6_{34}-5_{41}$&  --  & --   &  --  &  --  &  --  & --   &  --  &  --  & --\\
$3_{12}-2_{21}$& <0.1 & <0.1 &  --  & <0.1 &  --  & --   &  --  & <0.3 &<0.7\\
$1_{11}-0_{00}$& <0.1 & <0.1 &  --  & <0.1 &  --  & <0.3 &  --  & <0.7 & --\\
$3_{12}-3_{03}$& <0.4 & <0.1 &  --  & <0.2 &  --  & <0.1 & 0.5  &  --  & --\\
$2_{02}-1_{11}$& <0.2 & <0.1 &  --  & <0.1 &  --  & <0.1 &  --  & <0.3 & --\\
$5_{24}-4_{31}$&  --  & --   &  --  &  --  &  --  & --   &  --  &  --  & --\\
$4_{22}-3_{31}$&  --  & --   &  --  &  --  &  --  & --   &  --  &  --  & --\\
$2_{11}-2_{02}$& <0.3 & <0.1 & <0.3 & <0.2 & <0.5 & <0.3 & <0.2 & <0.5 & --\\
$5_{32}-4_{41}$&  --  & --   &  --  &  --  &  --  &  --  &  --  &  --  & --\\
$1_{10}-1_{01}$& <0.3 & <0.2 &  --  &  --  &  --  & <0.6 & <0.4 &  --  & --\\
\hline
\end{tabular}
\tablefoot{The "--" symbol indicates that both H$_{2}^{18}$O and 
H$_{2}^{17}$O are not detected.}
\end{table*}


\Online


\section{Compliation of observed SPIRE line fluxes}
\label{app_spire}

We present a table with line fluxes for the SPIRE spectra of extreme OH/IR 
stars. 

\begin{longtab}
\begin{longtable}{lcccccccccc}
\caption{Line fluxes of all lines detected in the observed SPIRE
range.}\\
\label{tab_obs_spire_all}
Molecules & $\lambda (\mu$m) &
\multicolumn{9}{c}{ Line fluxes ($\times 10^{-17}$ W m$^{-2}$)} \\
\cline{3-11}
& &  OH127& GL5379& OH21.5& OH26.5& OH30.7& OH30.1& OH32.0& OH32.8& OH42.3\\
\hline
\endfirsthead
\caption{continued.}\\
\hline\hline
Molecules & $\lambda (\mu$m) &
\multicolumn{9}{c}{ Line fluxes ($\times 10^{-17}$ W m$^{-2}$)} \\
\cline{3-11}
& &  OH127& GL5379& OH21.5& OH26.5& OH30.7& OH30.1& OH32.0& OH32.8& OH42.3\\
\hline
\endhead
\hline
\endfoot
H$_{2}$O~$6_{33}-5_{42}$      & 194.4221     & 2.2  & 22.5 & <0.3 & 7.9  & <0.7 & 1.7  & 1.0  & 1.7  & 0.7  \\
CO (13-12)                    & 200.2725     & 1.1  & 11.3 & <0.3 & 5.6  & <0.7 & 2.9  & 1.5  & 2.7  & 1.7  \\
H$_{2}$O~$\nu_{2}~2_{20}-2_{11}$ & 200.6566  & 0.5  & 8.9  & 1.3  & 3.7  & <0.7 & <0.3 & 1.7  &  1.8 & 0.9  \\
H$_{2}$O~$\nu_{2}~5_{23}-4_{32}$ & 203.4464  & 0.7  & 2.2  & <0.3 & <0.3 & <0.7 & <0.3 & <0.3 & <0.6 & <0.4 \\
$[\ion{N}{II}]$ $^{3}$P$_{1}-^{3}$P$_{0}$  & 205.1733  & <0.2 & 10.1 & 17.4 & 25.0 & 38.6 & 31.2 & 34.6 & 20.3  & 11.5 \\
H$_{2}$O~$7_{26}-6_{33}$      & 208.0763     & 1.0  & 14.1 & <0.3 & 5.2  & <0.7 & 1.3  & <0.3 & 0.9  & 0.7  \\
H$_{2}$O~$\nu_{2}~4_{22}-4_{13}$ & 210.8308  & 0.6  & 7.8  & <0.3 & 3.0  & <0.7 & <0.3 & 0.4  & 0.7 & <0.4 \\
H$_{2}$O~$5_{23}-5_{14}$      & 212.5256     & 4.1  & 57.9 & 1.8  & 14.3 & <0.7 & 9.0  & 2.0  & 5.3  & 2.3  \\
\hline
\end{longtable}
\end{longtab}

\clearpage

\section{Compliation of observed PACS line fluxes}
\label{app_linefluxes}

We present tables for the list of lines detected in the PACS spectra 
for individual objects.

\begin{longtab}
\begin{longtable}{lrc}
\caption{Line fluxes calculated from PACS spectra of AFGL~5379.}\\
\hline\hline
Molecules & $\lambda (\mu$m) & F$_{\rm line}$ ($\times 10^{-17}$ W m$^{-2}$) \\
\hline
\endfirsthead
\caption{continued.}\\
\hline\hline
Molecules & $\lambda (\mu$m) & F$_{\rm line}$ ($\times 10^{-17}$ W m$^{-2}$) \\
\hline
\endhead
\hline
\endfoot
H$_{2}$O         4$_{31}- 3_{22}$ &     56.3250 &  88.9 $\pm$ 35.6 \\
H$_{2}$O         4$_{22}- 3_{13}$ &     57.6365 & 155.2 $\pm$ 57.4 \\
H$_{2}$O         4$_{32}- 3_{21}$ &     58.6991 & 166.6 $\pm$ 38.7 \\
H$_{2}$O         8$_{26}- 7_{35}$ &     60.1622 &  78.5 $\pm$ 40.7 \\
H$_{2}$O         6$_{60}- 6_{51}$ &     63.9281 & 164.5 $\pm$ 45.9 \\
H$_{2}$O         9$_{64}- 9_{55}$ &     63.9940 &  81.6 $\pm$141.5 \\
H$_{2}$O         6$_{25}- 5_{14}$ &     65.1662 & 161.0 $\pm$ 14.7 \\
H$_{2}$O        10$_{38}-10_{29}$ &     65.5763 &  96.2 $\pm$ 14.9 \\
\end{longtable}
\tablefoot{
\tablefoottext{a}{the line is blended with a nearby transition}
}
\end{longtab}

\setcounter{table}{1}

\begin{longtab}
\begin{longtable}{lrc}
\caption{Line fluxes calculated from PACS spectra of OH~26.5+0.6.}\\
\hline\hline
Molecules & $\lambda (\mu$m) & F$_{\rm line}$ ($\times 10^{-17}$ W m$^{-2}$) \\
\hline
\endfirsthead
\caption{continued.}\\
\hline\hline
Molecules & $\lambda (\mu$m) & F$_{\rm line}$ ($\times 10^{-17}$ W m$^{-2}$) \\
\hline
\endhead
\hline
\endfoot
H$_{2}$O         8$_{27}- 7_{16}$ &     55.1310 & 108.2 $\pm$ 18.4 \\
H$_{2}$O         4$_{22}- 3_{13}$ &     57.6365 &  57.7 $\pm$ 14.4 \\
H$_{2}$O         4$_{32}- 3_{21}$ &     58.6991 &  76.1 $\pm$ 11.9 \\
H$_{2}$O         7$_{26}- 6_{15}$ &     59.9871 &  53.7 $\pm$  9.3 \\
H$_{2}$O         8$_{26}- 7_{35}$ &     60.1622 &  44.1 $\pm$  9.4 \\
H$_{2}$O         4$_{31}- 4_{04}$ &     61.8086 &  83.2 $\pm$ 13.2 \\
H$_{2}$O         4$_{40}- 5_{15}$ &     61.9162 &  38.0 $\pm$ 13.2 \\
H$_{2}$O         9$_{36}- 8_{45}$ &     62.4179 &  13.5 $\pm$ 13.5 \\
\end{longtable}
\tablefoot{
\tablefoottext{a}{the line is blended with a nearby transition}
}
\end{longtab}

\setcounter{table}{2}

\begin{longtab}
\begin{longtable}{lrc}
\caption{Line fluxes calculated from PACS spectra of OH~30.1-0.7.}\\
\hline\hline
Molecules & $\lambda (\mu$m) & F$_{\rm line}$ ($\times 10^{-17}$ W m$^{-2}$) \\
\hline
\endfirsthead
\caption{continued.}\\
\hline\hline
Molecules & $\lambda (\mu$m) & F$_{\rm line}$ ($\times 10^{-17}$ W m$^{-2}$) \\
\hline
\endhead
\hline
\endfoot
H$_{2}$O         4$_{22}- 3_{13}$ &     57.6365 &  10.6 $\pm$  2.9 \\
H$_{2}$O         4$_{32}- 3_{21}$ &     58.6991 &  16.8 $\pm$  3.0 \\
H$_{2}$O         4$_{31}- 4_{04}$ &     61.8086 &  18.6 $\pm$  3.3 \\
H$_{2}$O         6$_{60}- 6_{51}$ &     63.9281 &  18.5 $\pm$  3.5 \\
H$_{2}$O         6$_{25}- 5_{14}$ &     65.1662 &  15.4 $\pm$  3.3 \\
H$_{2}$O        10$_{38}-10_{29}$ &     65.5763 &   9.7 $\pm$  3.3 \\
H$_{2}$O         7$_{16}- 6_{25}$ &     66.0927 &  11.4 $\pm$  3.4 \\
H$_{2}$O         3$_{31}- 2_{20}$ &     67.0892 &  16.3 $\pm$  3.5 \\
\end{longtable}
\tablefoot{
\tablefoottext{a}{the line is blended with a nearby transition}
}
\end{longtab}

\setcounter{table}{3}

\begin{longtab}
\begin{longtable}{lrc}
\caption{Line fluxes calculated from PACS spectra of OH~32.0-0.5.}\\
\hline\hline
Molecules & $\lambda (\mu$m) & F$_{\rm line}$ ($\times 10^{-17}$ W m$^{-2}$) \\
\hline
\endfirsthead
\caption{continued.}\\
\hline\hline
Molecules & $\lambda (\mu$m) & F$_{\rm line}$ ($\times 10^{-17}$ W m$^{-2}$) \\
\hline
\endhead
\hline
\endfoot
H$_{2}$O         6$_{60}- 6_{51}$ &     63.9281 &  10.3 $\pm$ 28.2 \\
H$_{2}$O         6$_{25}- 5_{14}$ &     65.1662 &   4.8 $\pm$  2.9 \\
H$_{2}$O         3$_{30}- 2_{21}$ &     66.4377 &   4.3 $\pm$  3.1 \\
H$_{2}$O         3$_{31}- 2_{20}$ &     67.0892 &   5.7 $\pm$  1.9 \\
H$_{2}$O         3$_{30}- 3_{03}$ &     67.2690 &   7.7 $\pm$  2.0 \\
H$_{2}$O $\nu_{2}$   3$_{21}- 2_{12}$ &     70.2871 &   3.5 $\pm$  2.1 \\
H$_{2}$O         8$_{27}- 8_{18}$ &     70.7026 &   6.6 $\pm$  2.6 \\
H$_{2}$O         5$_{24}- 4_{13}$ &     71.0673 &   5.7 $\pm$  2.6 \\
\end{longtable}
\tablefoot{
\tablefoottext{a}{the line is blended with a nearby transition}
}
\end{longtab}

\setcounter{table}{4}

\begin{longtab}
\begin{longtable}{lrc}
\caption{Line fluxes calculated from PACS spectra of OH~32.8-0.3.}\\
\hline\hline
Molecules & $\lambda (\mu$m) & F$_{\rm line}$ ($\times 10^{-17}$ W m$^{-2}$) \\
\hline
\endfirsthead
\caption{continued.}\\
\hline\hline
Molecules & $\lambda (\mu$m) & F$_{\rm line}$ ($\times 10^{-17}$ W m$^{-2}$) \\
\hline
\endhead
\hline
\endfoot
H$_{2}$O         4$_{32}- 3_{21}$ &     58.6991 &  17.4 $\pm$  3.9 \\
H$_{2}$O         7$_{26}- 6_{15}$ &     59.9871 &   9.6 $\pm$  4.0 \\
H$_{2}$O         8$_{26}- 7_{35}$ &     60.1622 &   9.1 $\pm$  4.1 \\
H$_{2}$O         4$_{31}- 4_{04}$ &     61.8086 &  15.2 $\pm$  4.3 \\
H$_{2}$O         9$_{28}- 9_{19}$ &     62.4316 &   8.8 $\pm$  4.4 \\
H$_{2}$O         8$_{18}- 7_{07}$ &     63.3236 &   6.7 $\pm$  4.5 \\
H$_{2}$O         8$_{08}- 7_{17}$ &     63.4580 &   7.5 $\pm$  4.5 \\
H$_{2}$O         6$_{60}- 6_{51}$ &     63.9281 &  26.0 $\pm$  4.6 \\
\end{longtable}
\tablefoot{
\tablefoottext{a}{the line is blended with a nearby transition}
}
\end{longtab}

\setcounter{table}{5}

\begin{longtab}
\begin{longtable}{lrc}
\caption{Line fluxes calculated from PACS spectra of OH~42.3-0.1.}\\
\hline\hline
Molecules & $\lambda (\mu$m) & F$_{\rm line}$ ($\times 10^{-17}$ W m$^{-2}$) \\
\hline
\endfirsthead
\caption{continued.}\\
\hline\hline
Molecules & $\lambda (\mu$m) & F$_{\rm line}$ ($\times 10^{-17}$ W m$^{-2}$) \\
\hline
\endhead
\hline
\endfoot
 H$_{2}$O         6$_{25}- 5_{14}$ &     65.1662 &   3.1 $\pm$  1.7 \\
 H$_{2}$O         3$_{30}- 2_{21}$ &     66.4377 &   3.1 $\pm$  1.7 \\
 H$_{2}$O         3$_{31}- 2_{20}$ &     67.0892 &   4.6 $\pm$  1.7 \\
 H$_{2}$O         3$_{30}- 3_{03}$ &     67.2690 &   6.9 $\pm$  1.8 \\
 H$_{2}$O         7$_{07}- 6_{16}$ &     71.9470 &   7.9 $\pm$  3.8 \\
 H$_{2}$O         9$_{37}- 9_{28}$ &     73.6129 &   5.7 $\pm$  2.8 \\
 H$_{2}$O $\nu_{2}$   4$_{23}- 3_{12}$ &     73.7448 &   6.4 $\pm$  2.8 \\
 H$_{2}$O        10$_{47}-10_{38}$ &     73.9603 &   4.9 $\pm$  2.8 \\
\end{longtable}
\tablefoot{
\tablefoottext{a}{the line is blended with a nearby transition}
}
\end{longtab}

\newpage

\section{SPIRE spectra}
\label{app_spire_plots}

This section shows the continuum subtracted apodized SPIRE
spectra (in W m$^{-2} \mu$m$^{-1}$) of the stars in our sample 
(histogram) together with
the Gaussian fits for H$_{2}$O (red),  H$_{2}^{17}$O (blue),
CO (green) and H$_{2}$S (yellow). Other molecules are shown in cyan.

\begin{figure*}
\centering
\includegraphics[width=17cm]{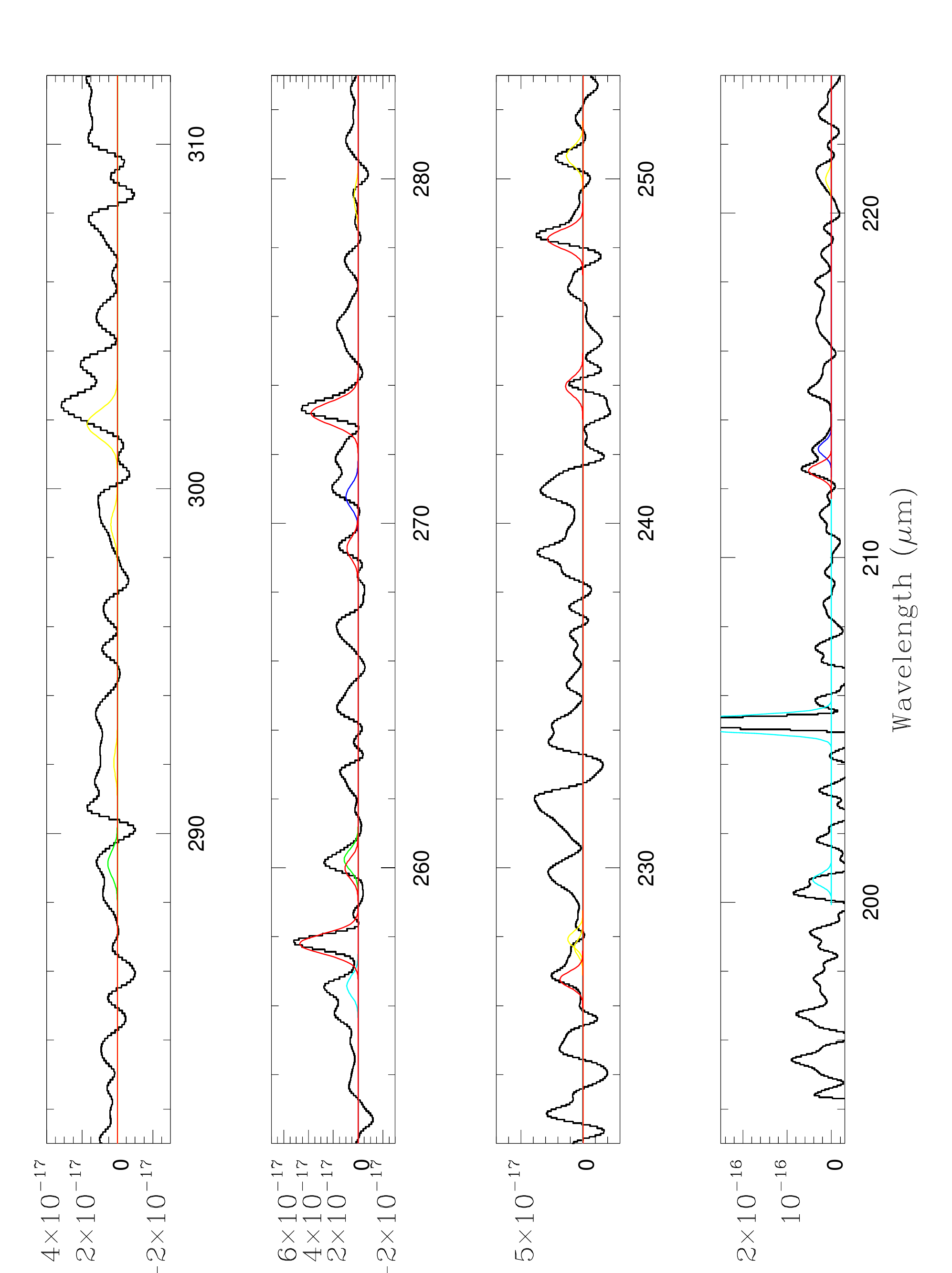}
\caption{The continuum subtracted apodized SPIRE spectrum of
OH21.5+0.5.
}
\end{figure*}

\begin{figure*}
\centering
\includegraphics[width=17cm]{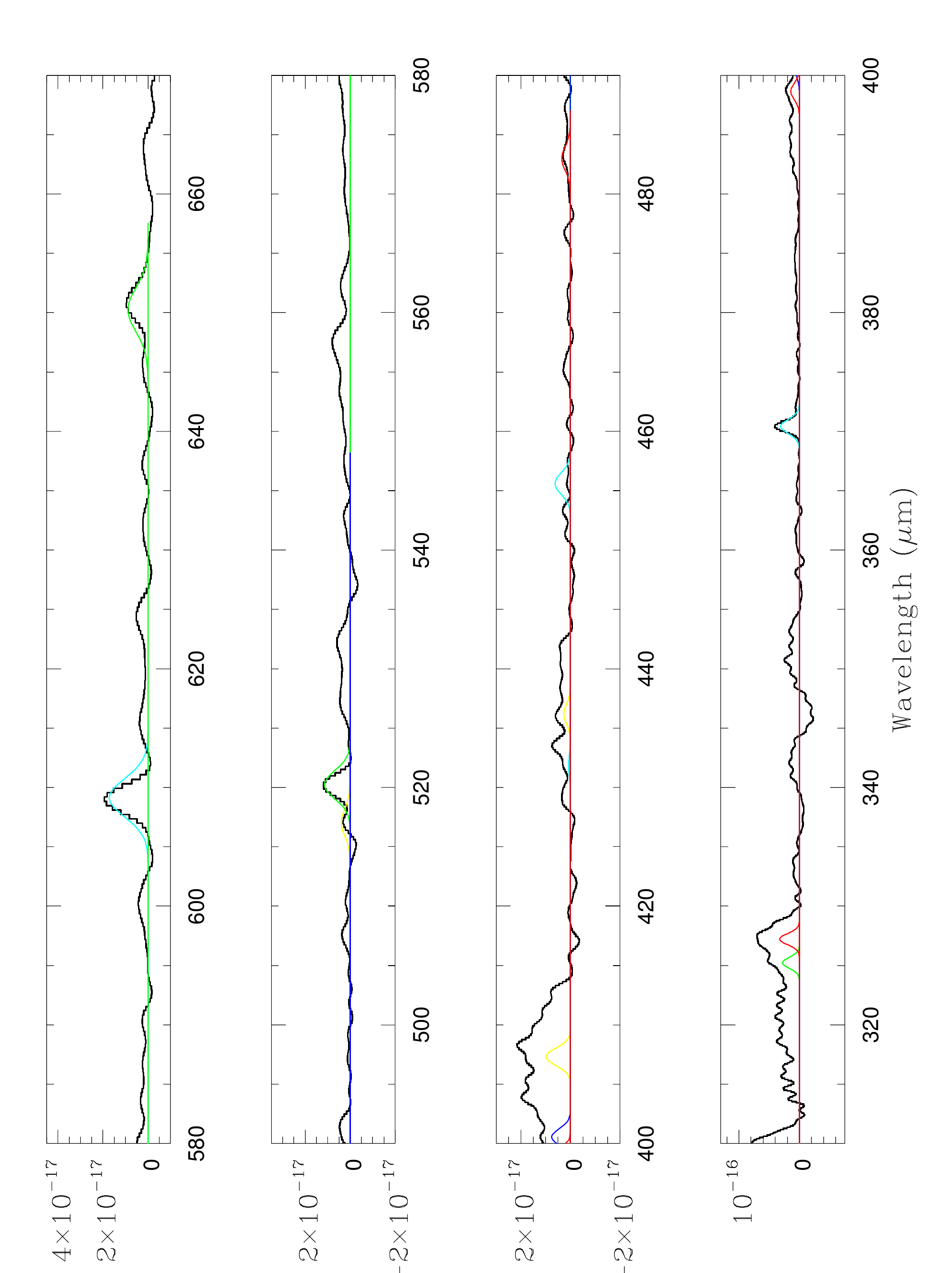}
\caption{The continuum subtracted apodized SPIRE spectrum of
OH21.5+0.5. 
}
\end{figure*}

\begin{figure*}
\centering
\includegraphics[width=17cm]{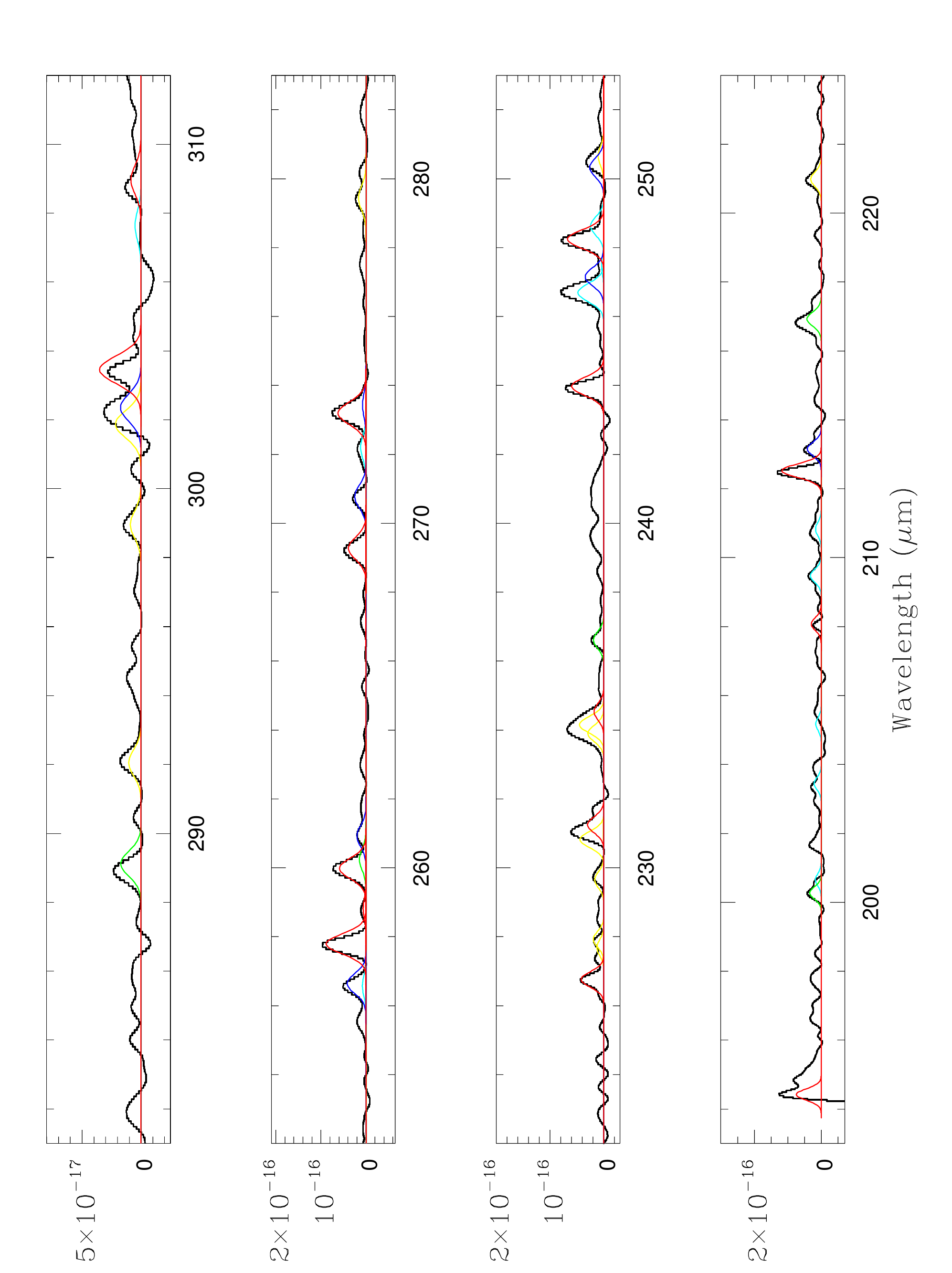}
\caption{The continuum subtracted apodized  SPIRE spectrum of
OH~127.8+0.0.}
\end{figure*}

\begin{figure*}
\centering
\includegraphics[width=17cm]{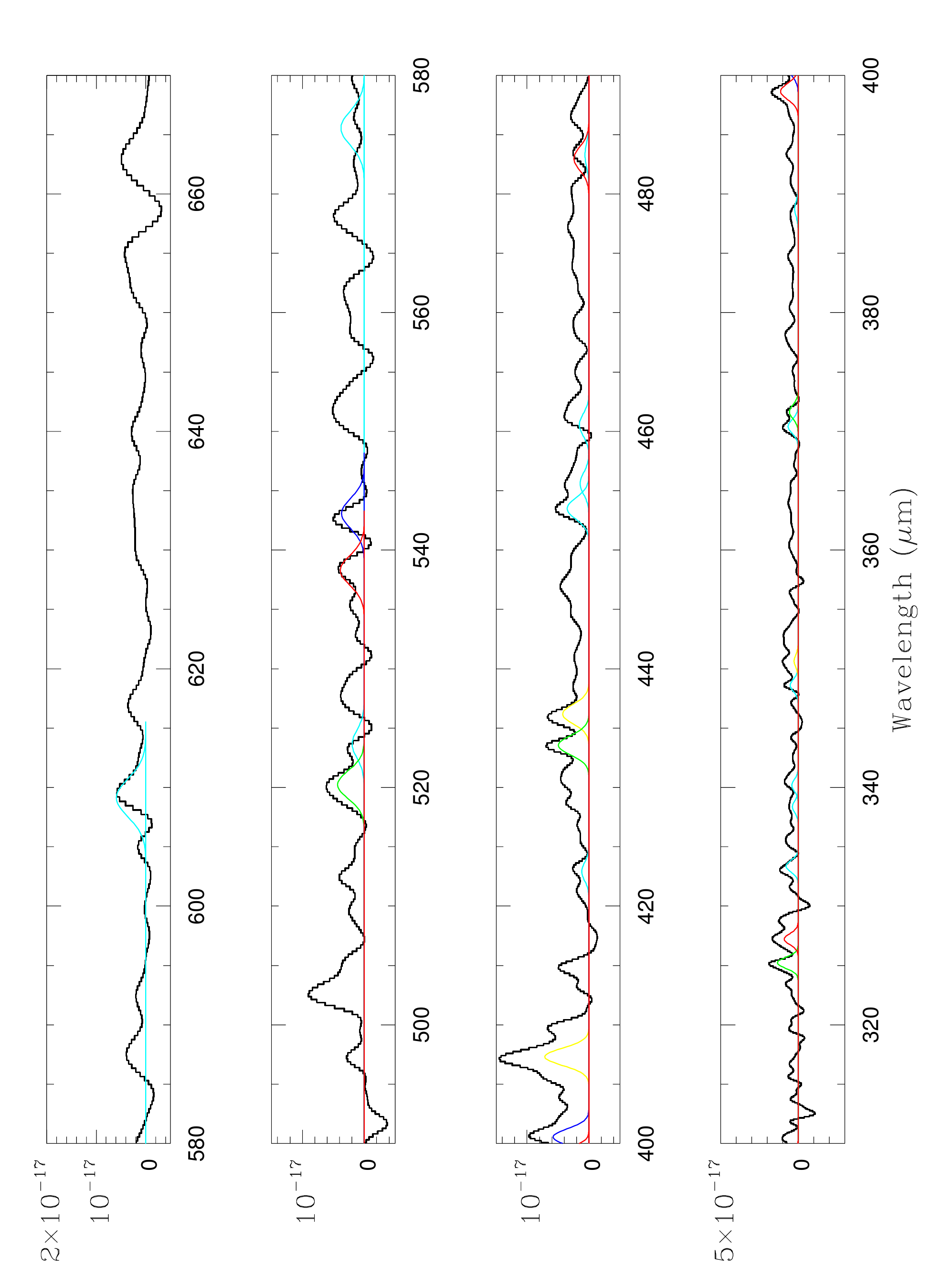}
\caption{The continuum subtracted apodized SPIRE spectrum of
OH~127.8+0.0.}
\end{figure*}

\begin{figure*}
\centering
\includegraphics[width=17cm]{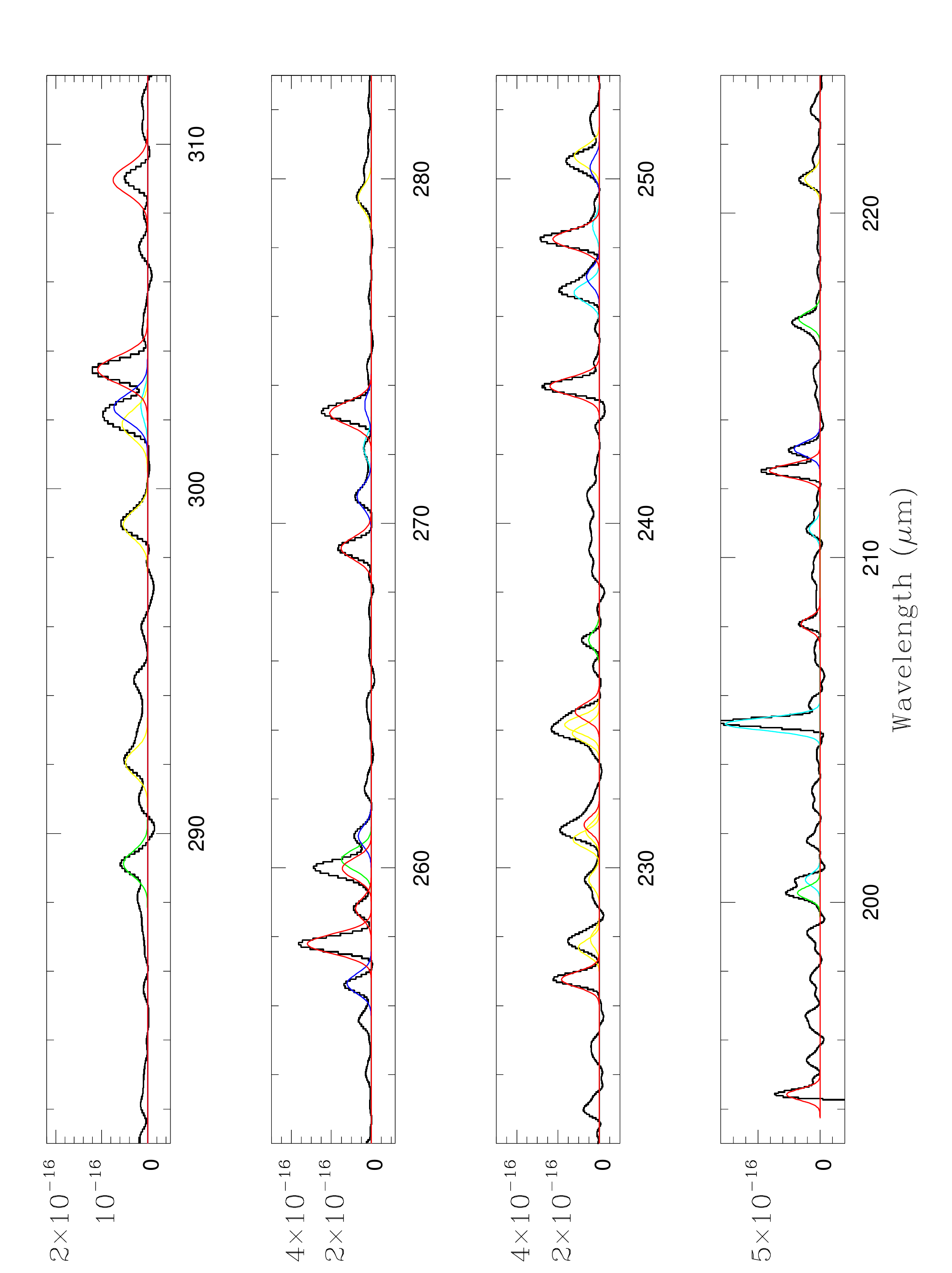}
\caption{The continuum subtracted apodized SPIRE spectrum of
OH~26.5+0.6.}
\end{figure*}

\begin{figure*}
\centering
\includegraphics[width=17cm]{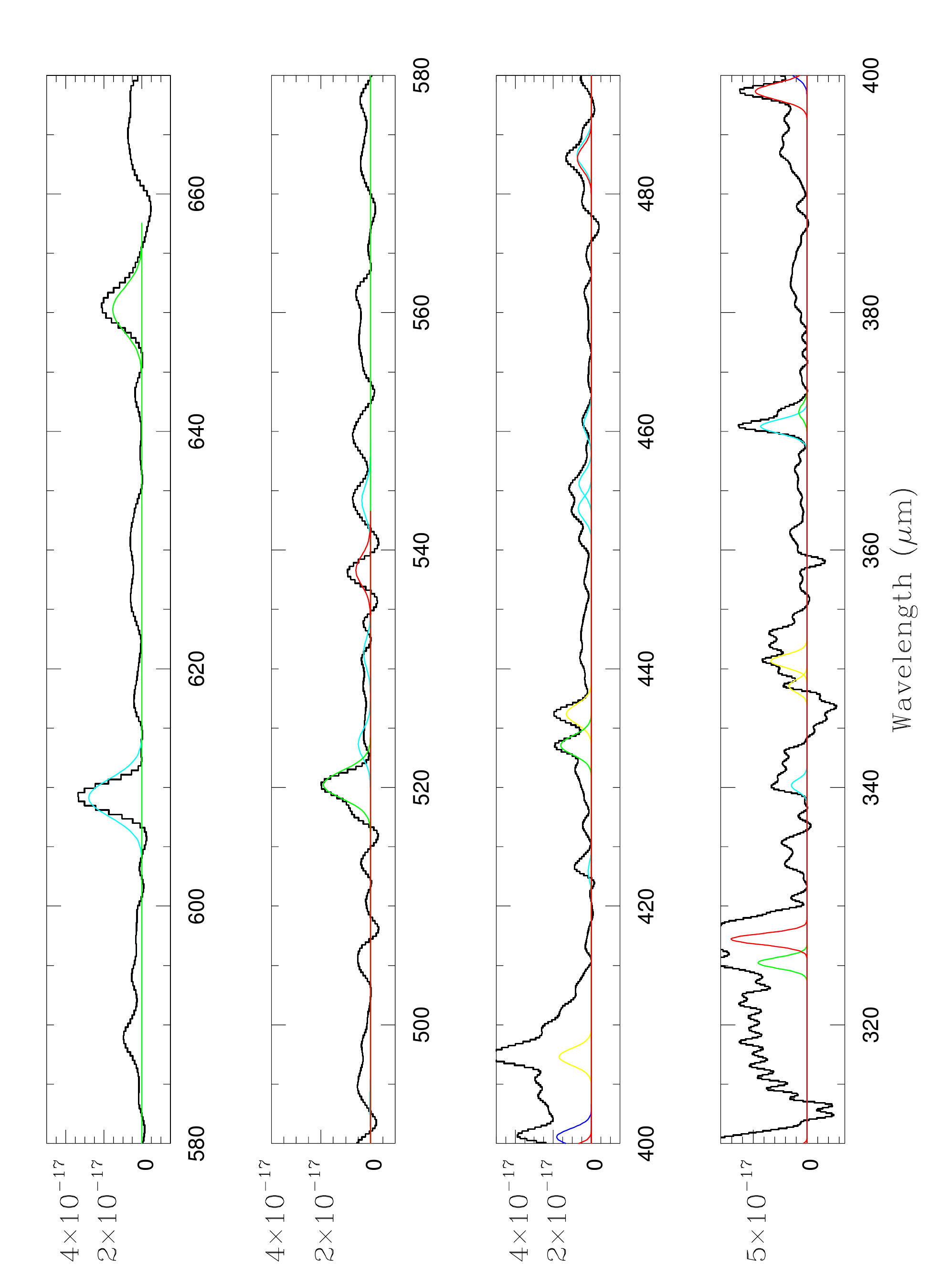}
\caption{The continuum subtracted apodized  SPIRE spectrum of
OH~26.5+0.6,}
\end{figure*}

\clearpage

\begin{figure*}
\centering
\includegraphics[width=17cm]{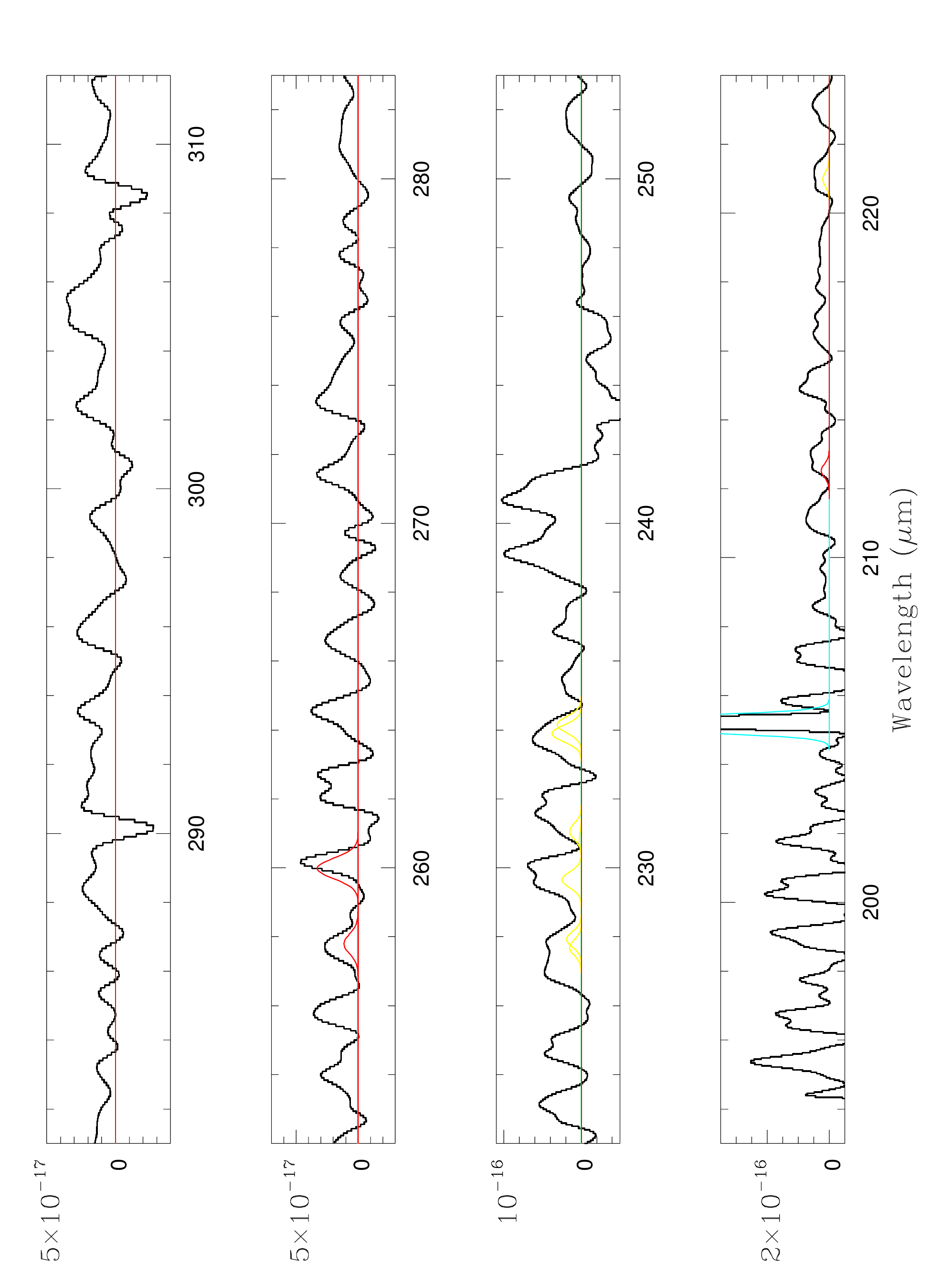}
\caption{The continuum subtracted apodized SPIRE spectrum of
OH~30.7+0.4.}
\end{figure*}

\begin{figure*}
\centering
\includegraphics[width=17cm]{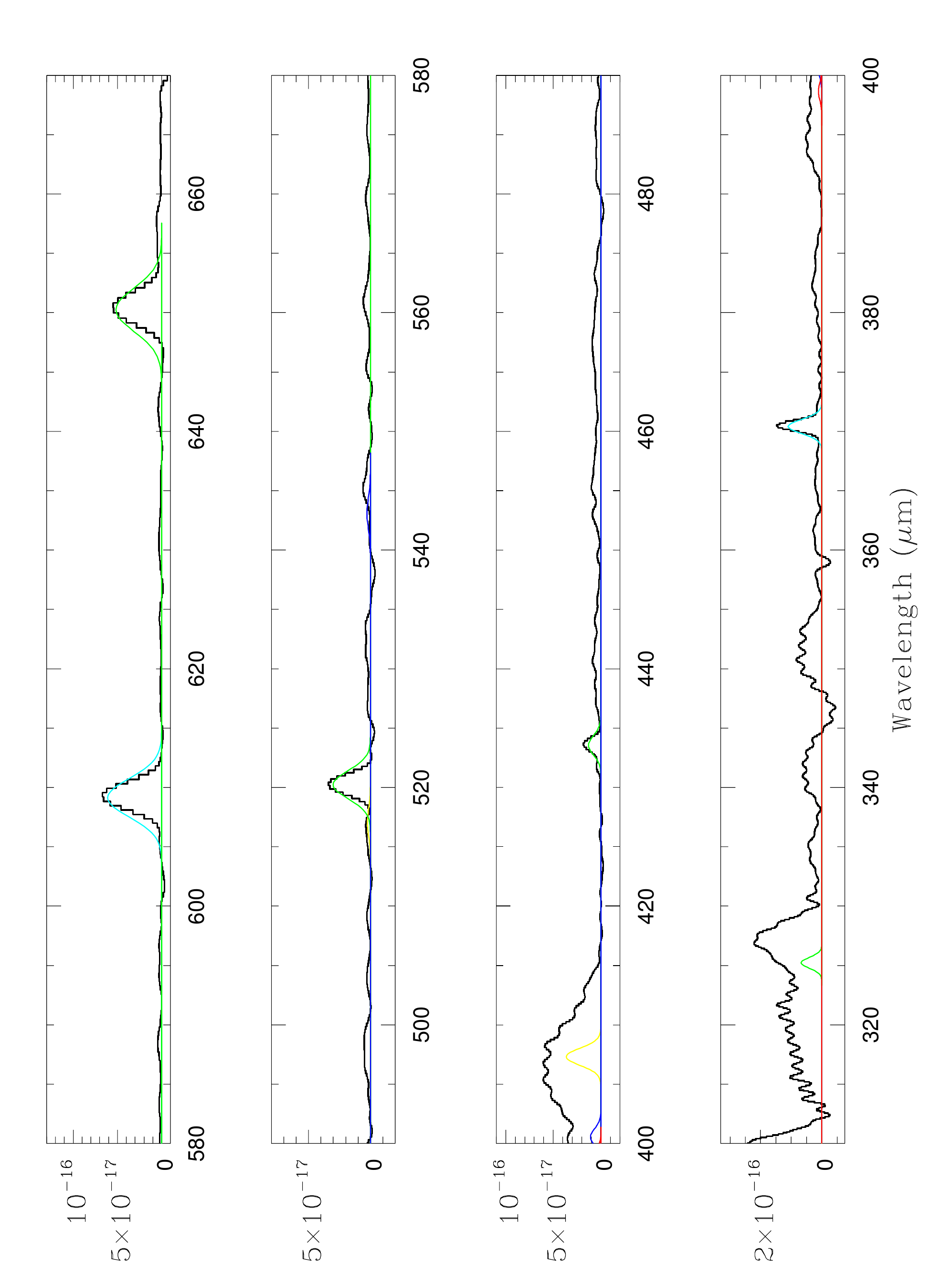}
\caption{The continuum subtracted apodized  SPIRE spectrum of
OH~30.7+0.4.}
\end{figure*}

\begin{figure*}
\centering
\includegraphics[width=17cm]{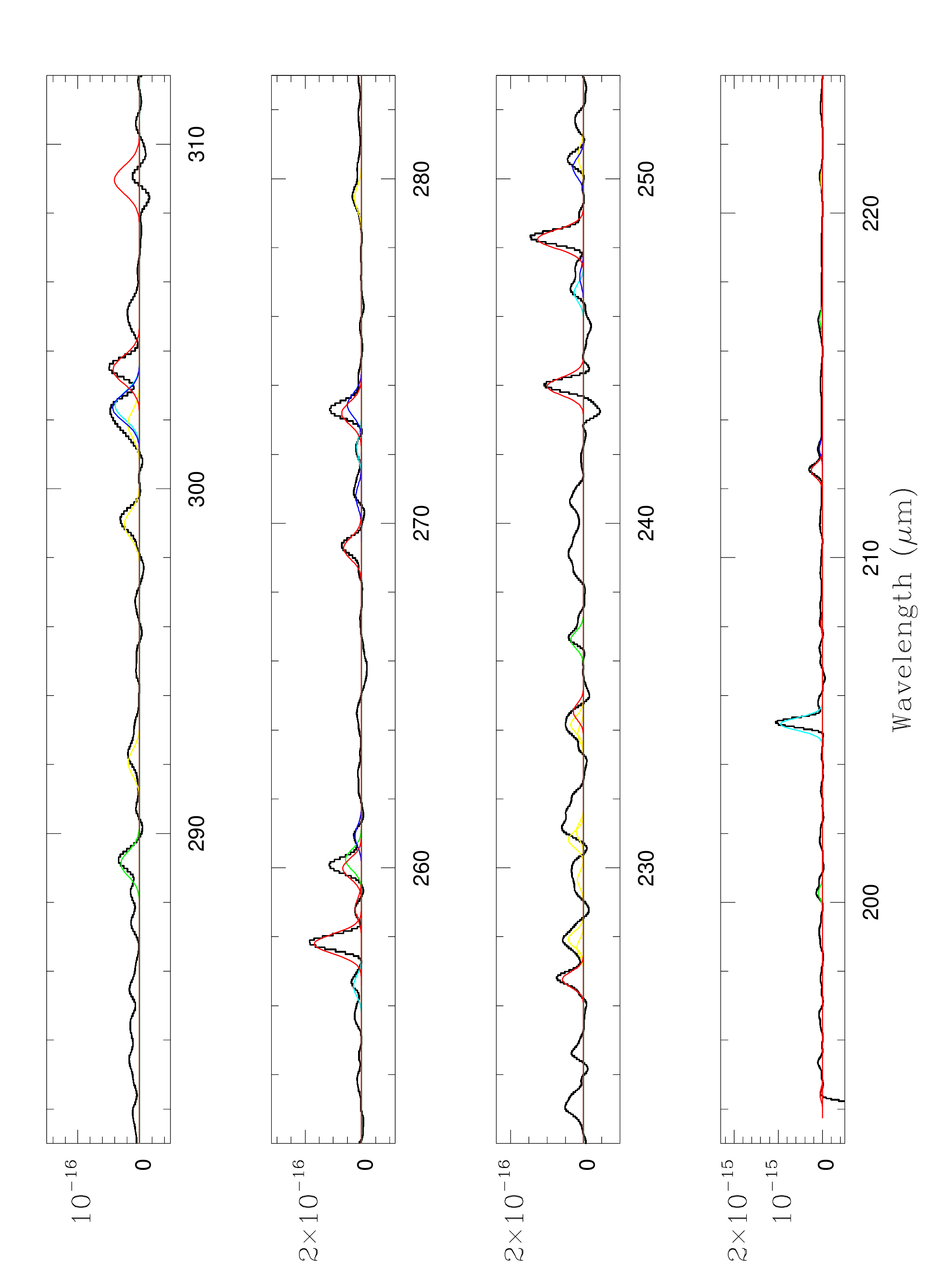}
\caption{The continuum subtracted apodized SPIRE spectrum of
OH~30.1-0.7.}
\end{figure*}

\begin{figure*}
\centering
\includegraphics[width=17cm]{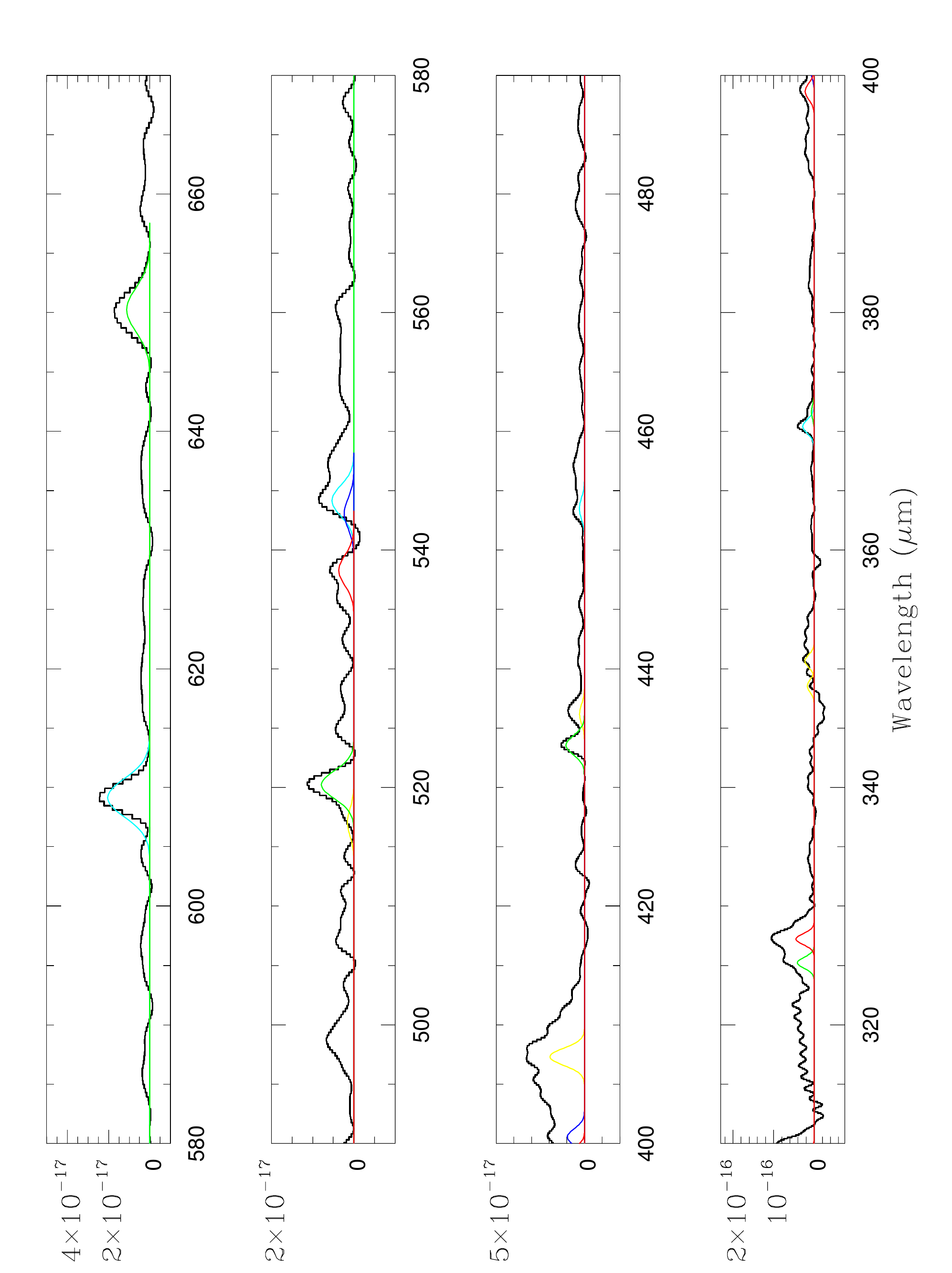}
\caption{The continuum subtracted apodized  SPIRE spectrum of
OH~30.1-0.7.}
\end{figure*}

\begin{figure*}
\centering
\includegraphics[width=17cm]{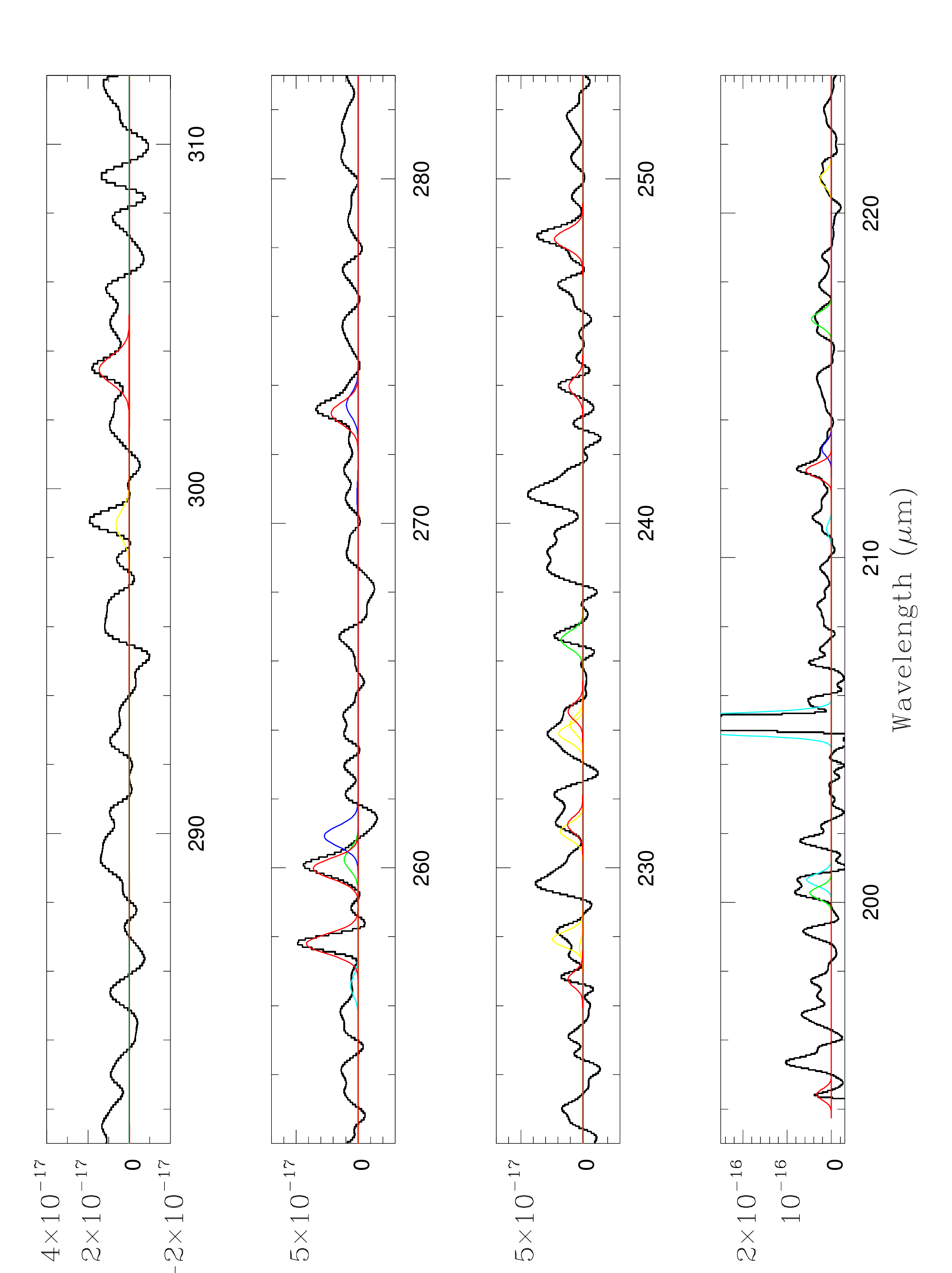}
\caption{The continuum subtracted apodized SPIRE spectrum of
OH~32.0-0.5.}
\end{figure*}

\begin{figure*}
\centering
\includegraphics[width=17cm]{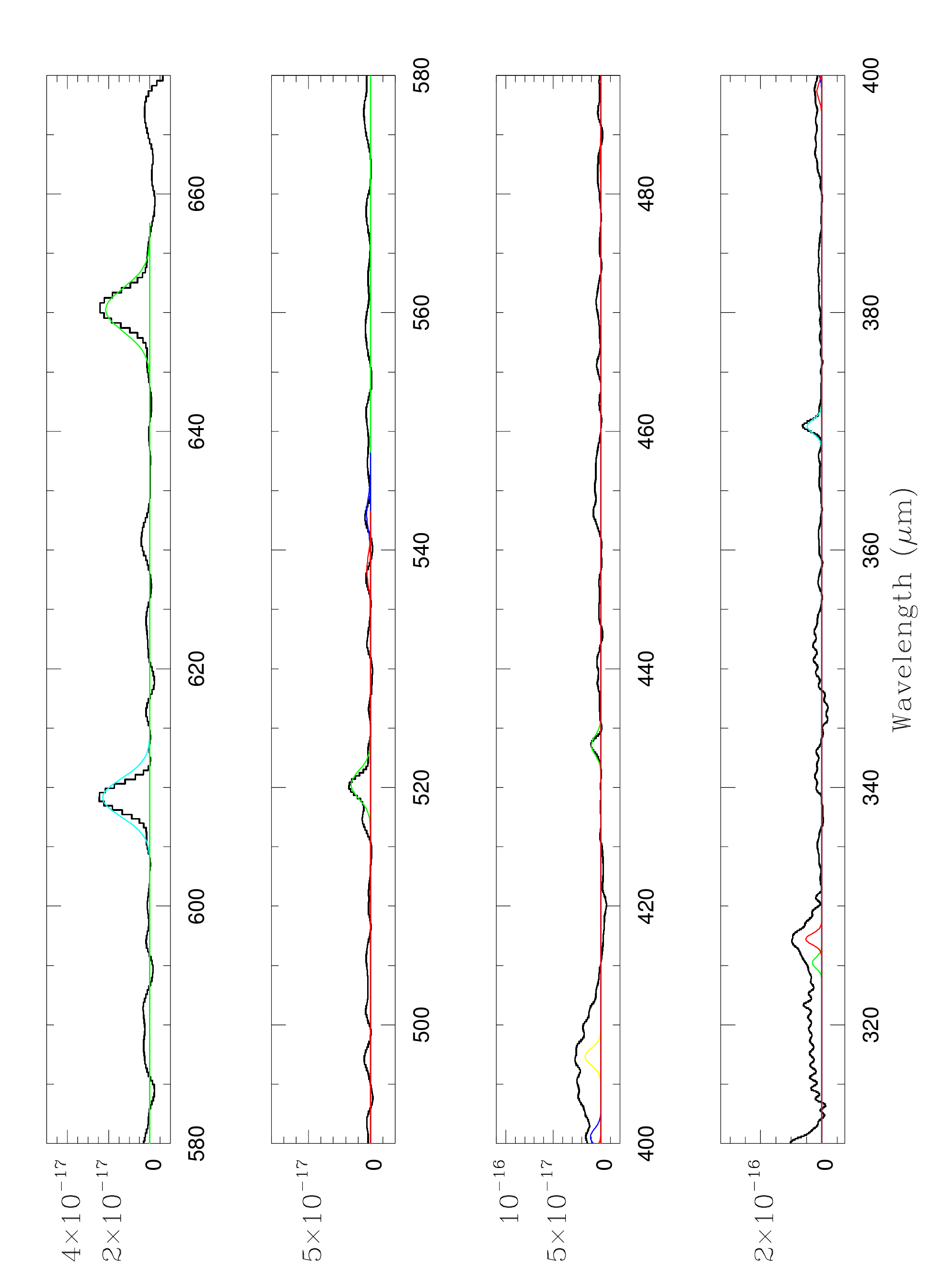}
\caption{The continuum subtracted apodized  SPIRE spectrum of
OH~32.0-0.5.}
\end{figure*}

\begin{figure*}
\centering
\includegraphics[width=17cm]{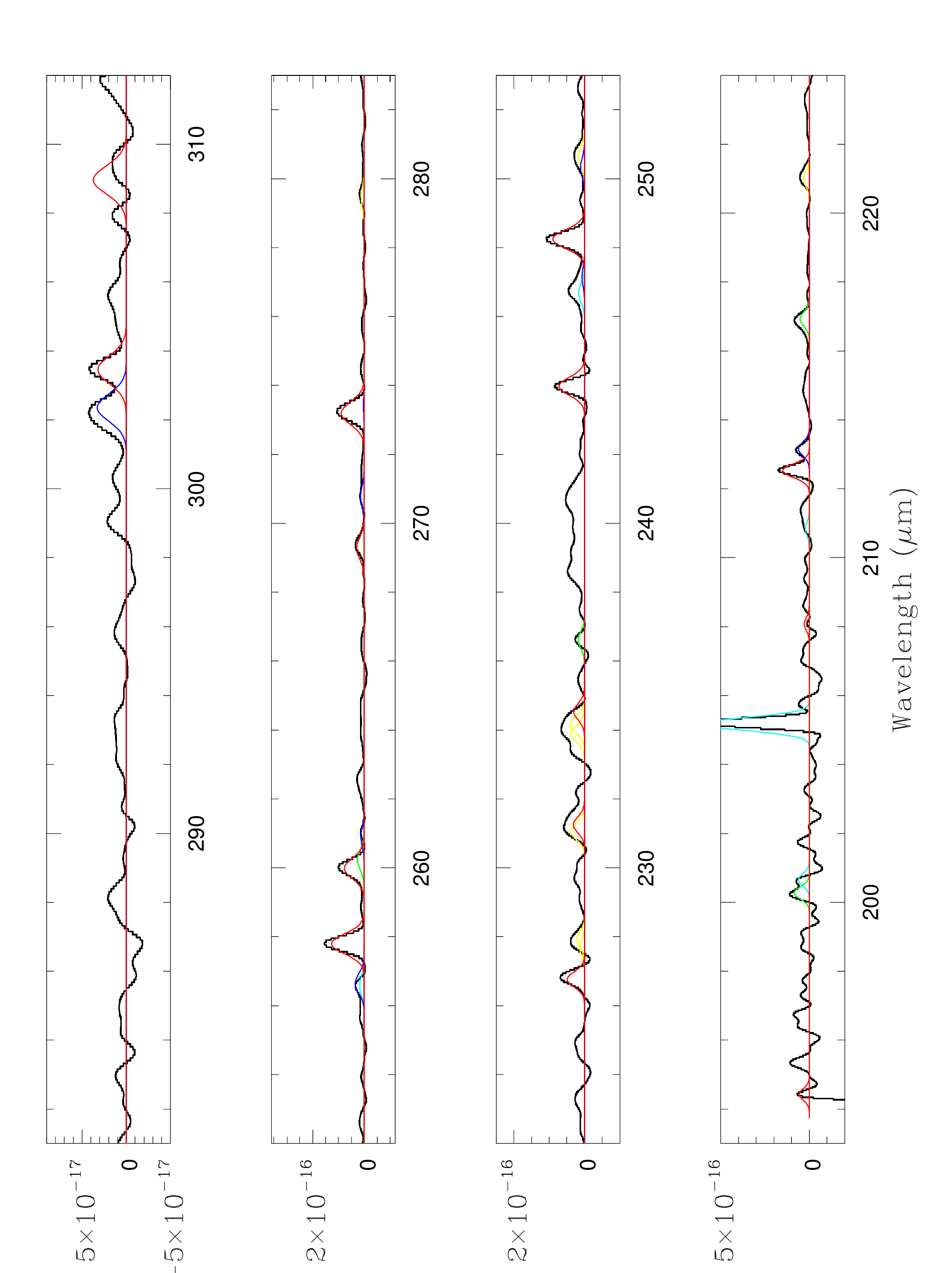}
\caption{The continuum subtracted apodized SPIRE spectrum of
OH~32.8-0.3.}
\end{figure*}

\begin{figure*}
\centering
\includegraphics[width=17cm]{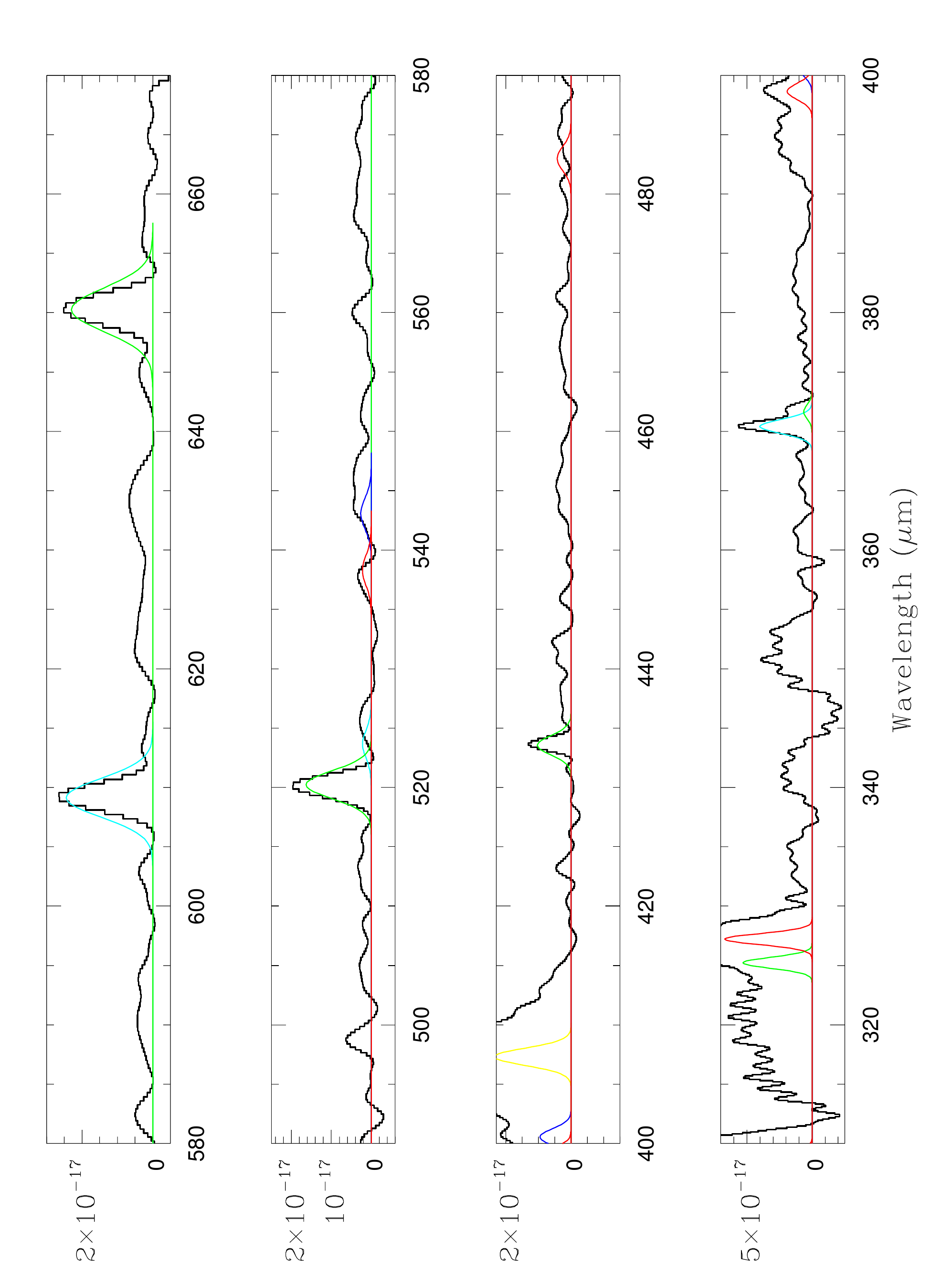}
\caption{The continuum subtracted apodized  SPIRE spectrum of
OH~32.8-0.3.}
\end{figure*}

\begin{figure*}
\centering
\includegraphics[width=17cm]{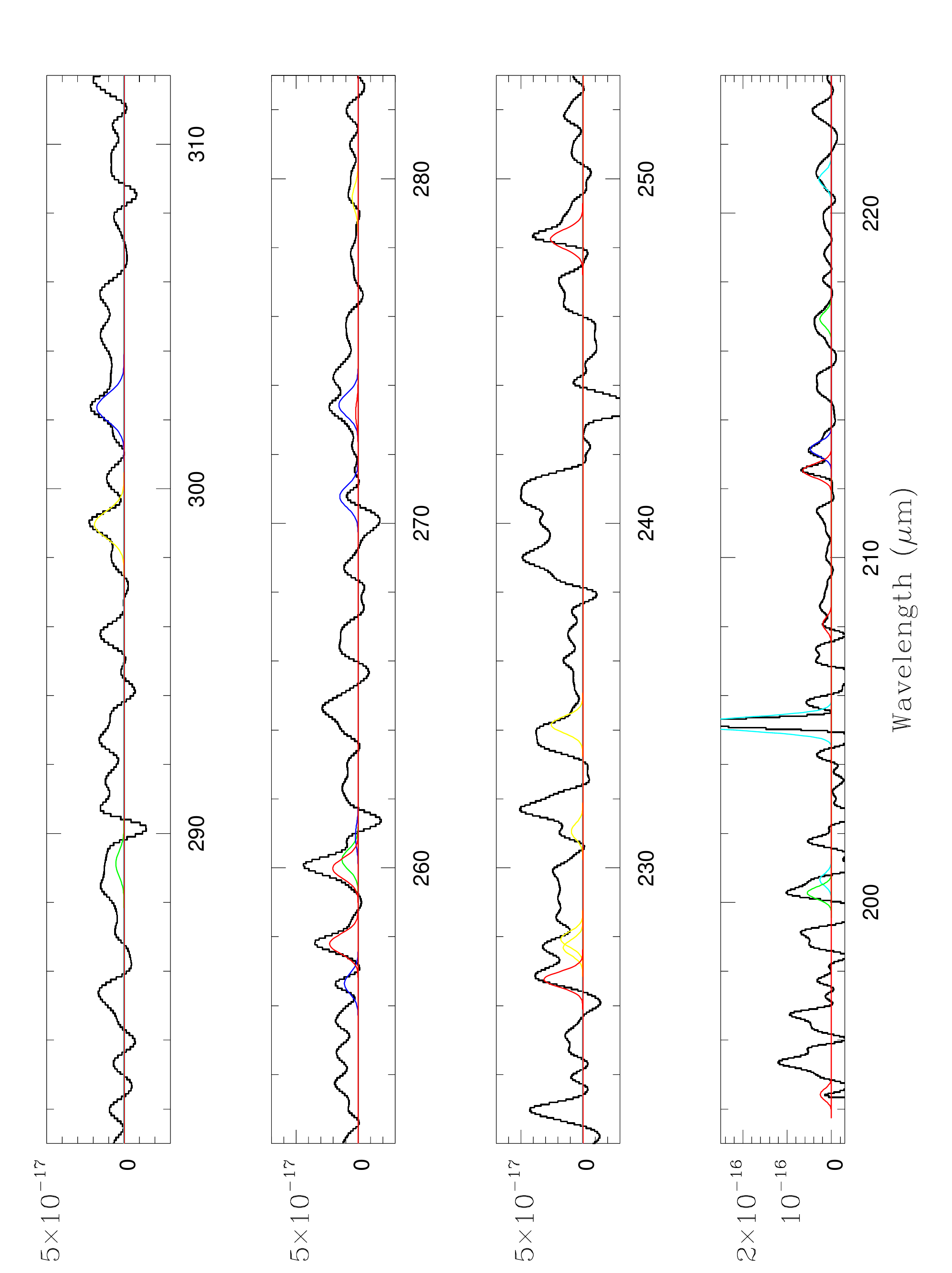}
\caption{The continuum subtracted apodized SPIRE spectrum of
OH~42.3-0.1.}
\end{figure*}

\begin{figure*}
\centering
\includegraphics[width=17cm]{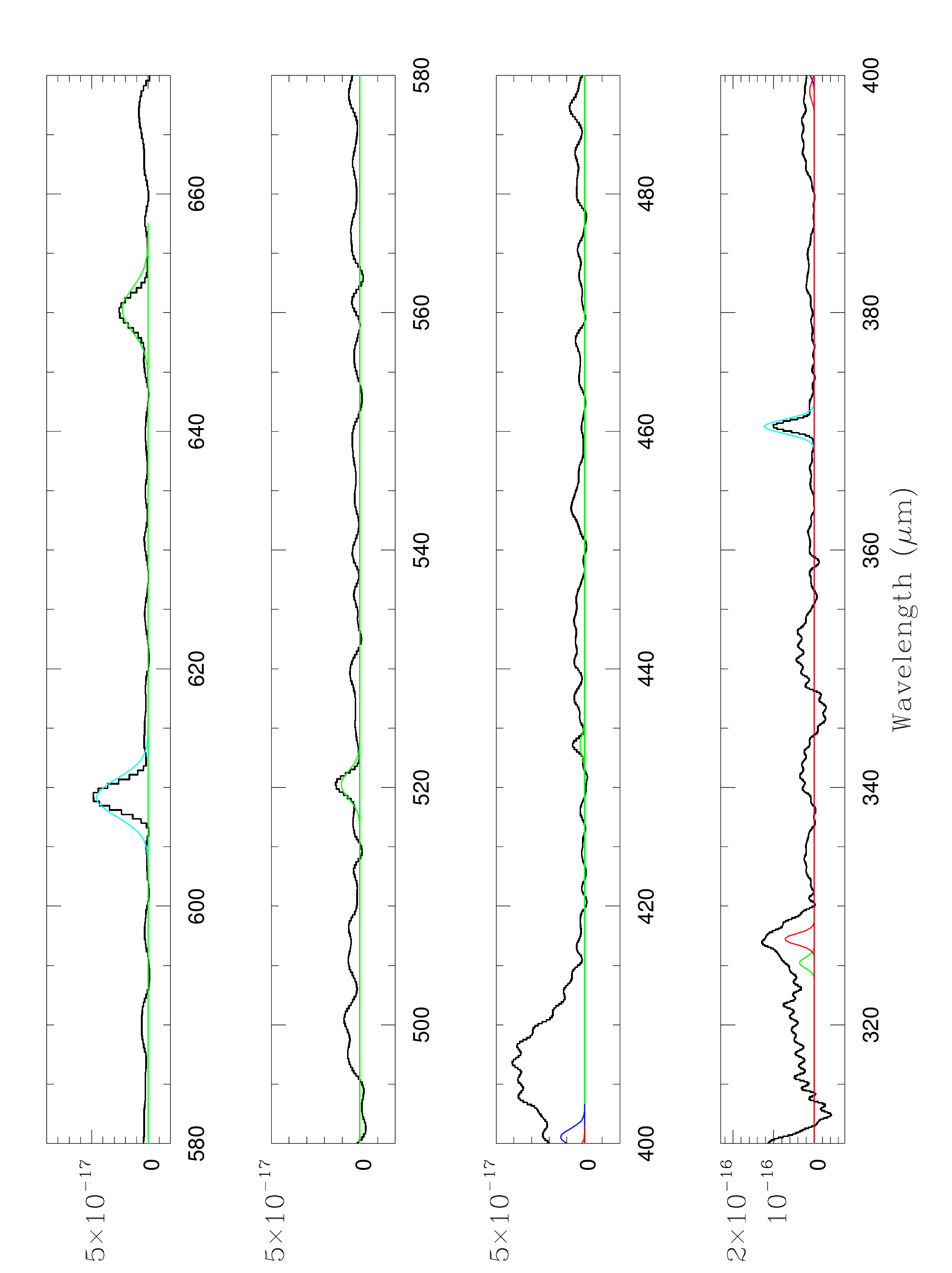}
\caption{The continuum subtracted apodized  SPIRE spectrum of
OH~42.3-0.1.}
\end{figure*}

\newpage

\section{PACS spectra}
\label{app_pacs_plots}

This section shows the continuum subtracted PACS
spectra (in W m$^{-2} \mu$m$^{-1}$) of the stars 
in our sample (histogram) together with
the Gaussian fits for H$_{2}$O (red),  H$_{2}^{17}$O (blue), and
CO (green). Other molecules are shown in cyan. We note that no
circumstellar emission lines are detected in the spectrum of
OH~21.5+0.5. The PACS spectrum of OH~127.8+0.0 has already been
publish by \cite{lombaert13} and is not presented here.

\clearpage
\begin{figure*}
\centering
\includegraphics[width=17cm]{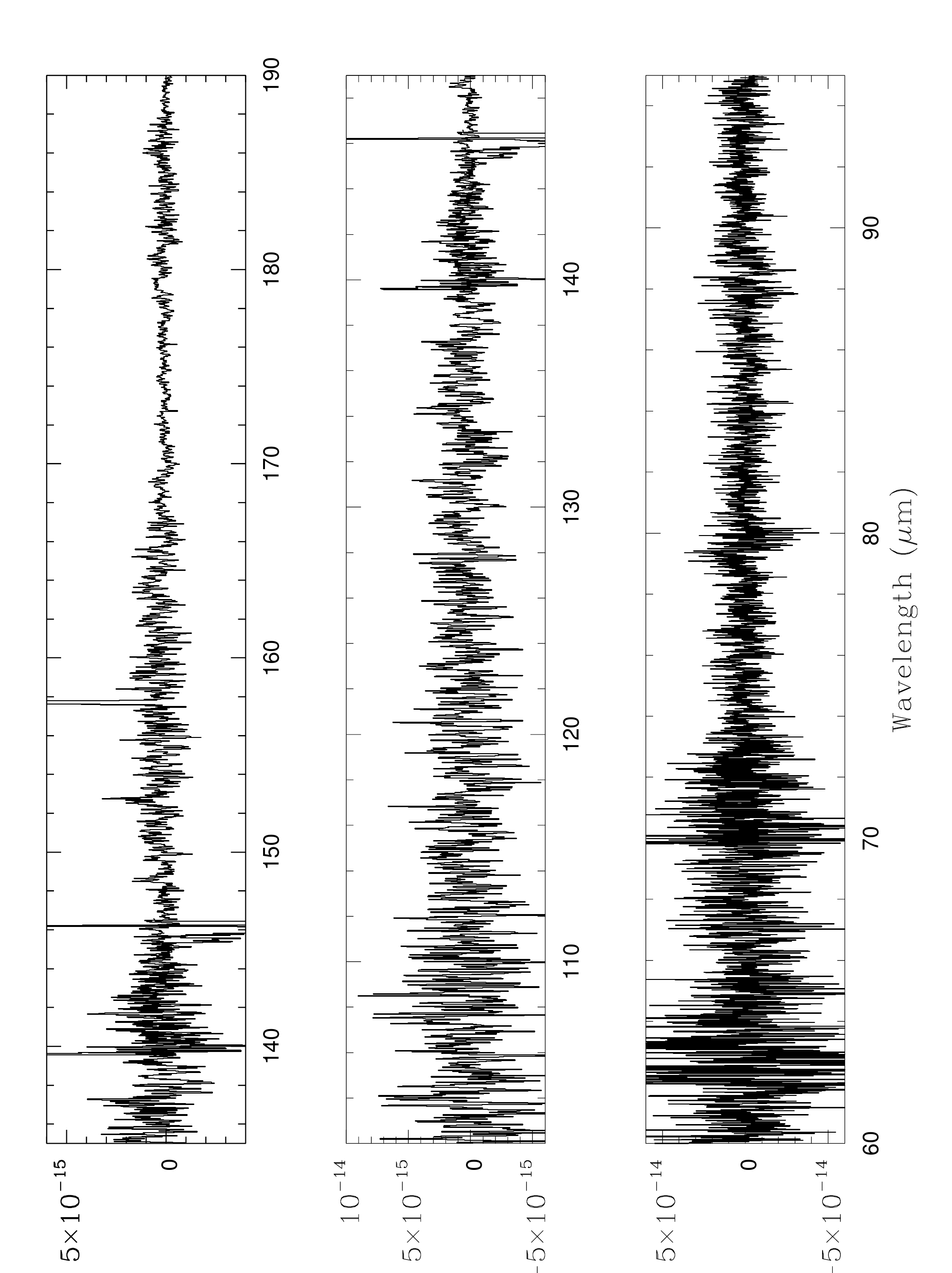}
\caption{The continuum subtracted PACS spectrum of OH~21.5+0.5.}
\end{figure*}

\begin{figure*}
\centering
\includegraphics[width=17cm]{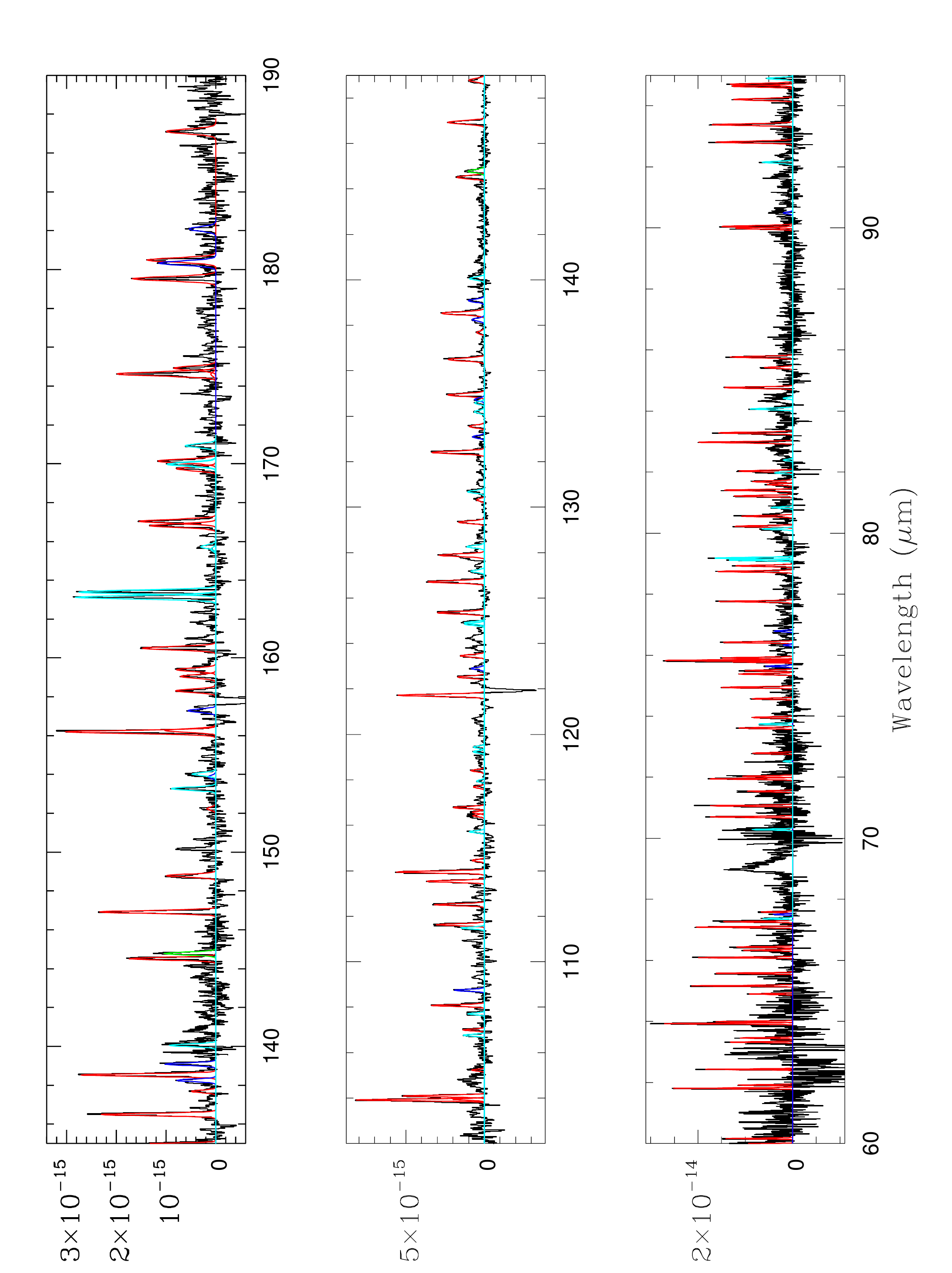}
\caption{The continuum subtracted PACS spectrum of OH~26.5+0.6.}
\end{figure*}

\begin{figure*}
\centering
\includegraphics[width=17cm]{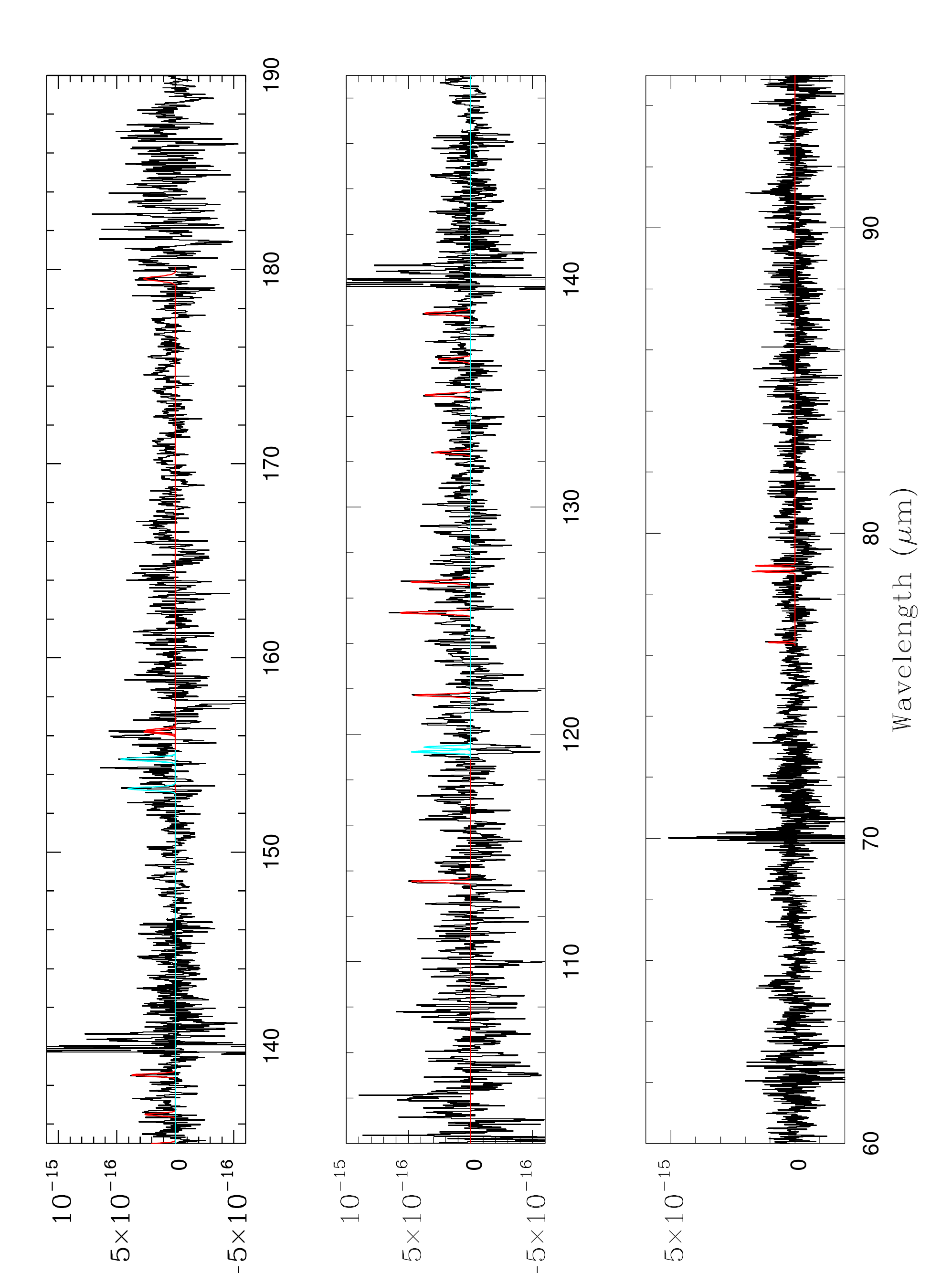}
\caption{The continuum subtracted PACS spectrum of OH~30.7+0.4.}
\end{figure*}

\begin{figure*}
\centering
\includegraphics[width=17cm]{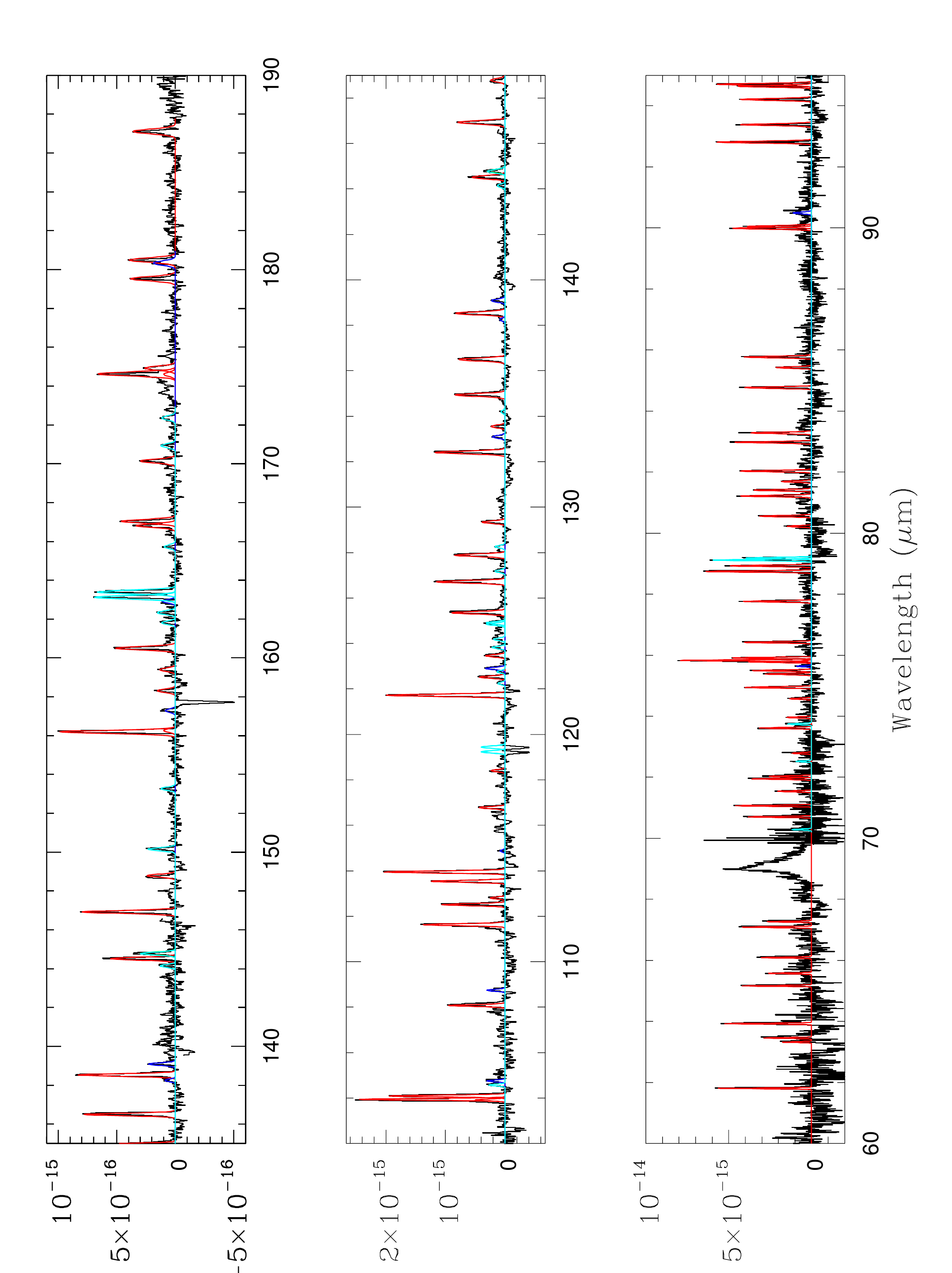}
\caption{The continuum subtracted PACS spectrum of OH~30.1-0.7.}
\end{figure*}

\begin{figure*}
\centering
\includegraphics[width=17cm]{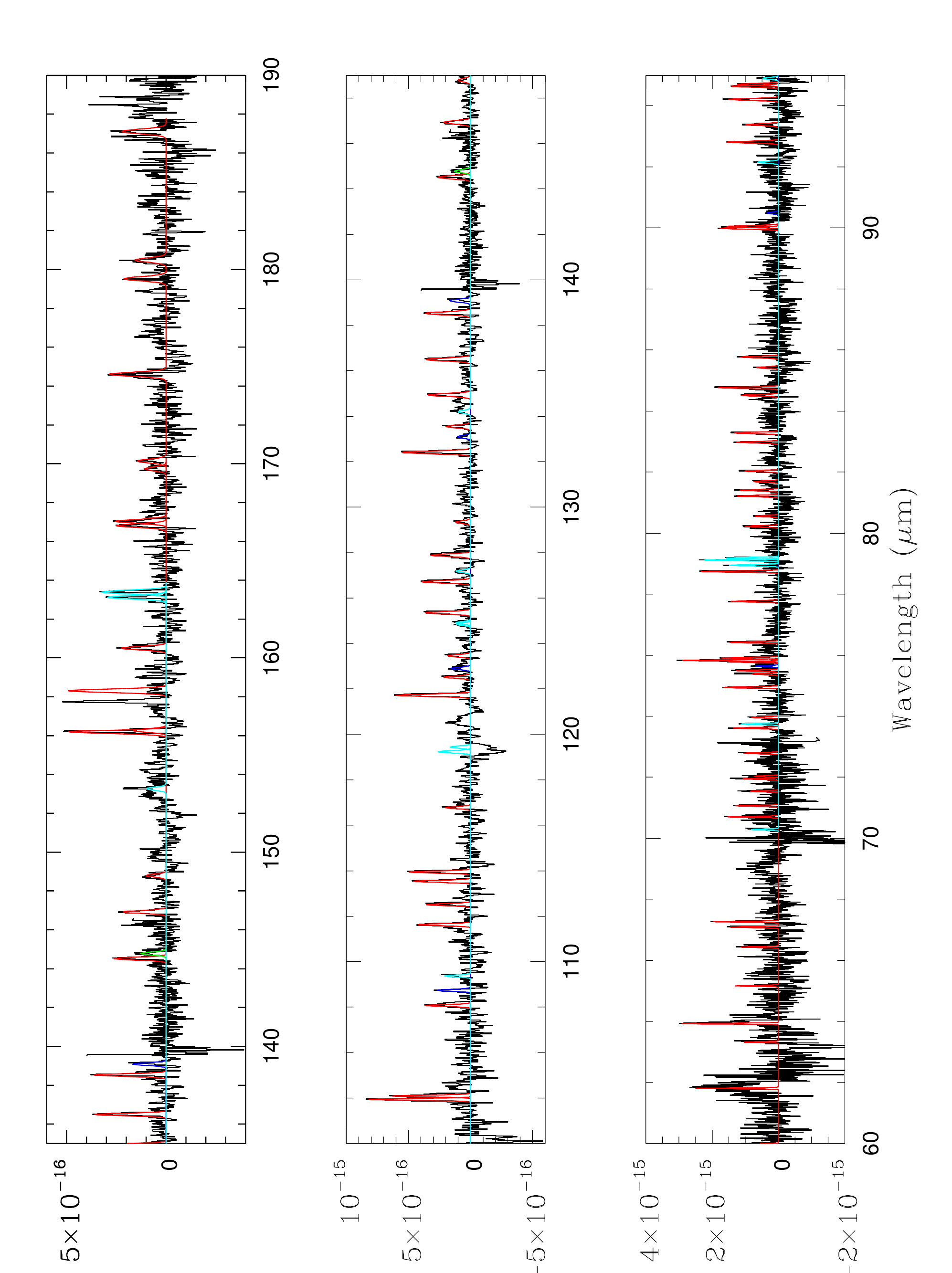}
\caption{The continuum subtracted PACS spectrum of OH~32.0-0.5.}
\end{figure*}

\begin{figure*}
\centering
\includegraphics[width=17cm]{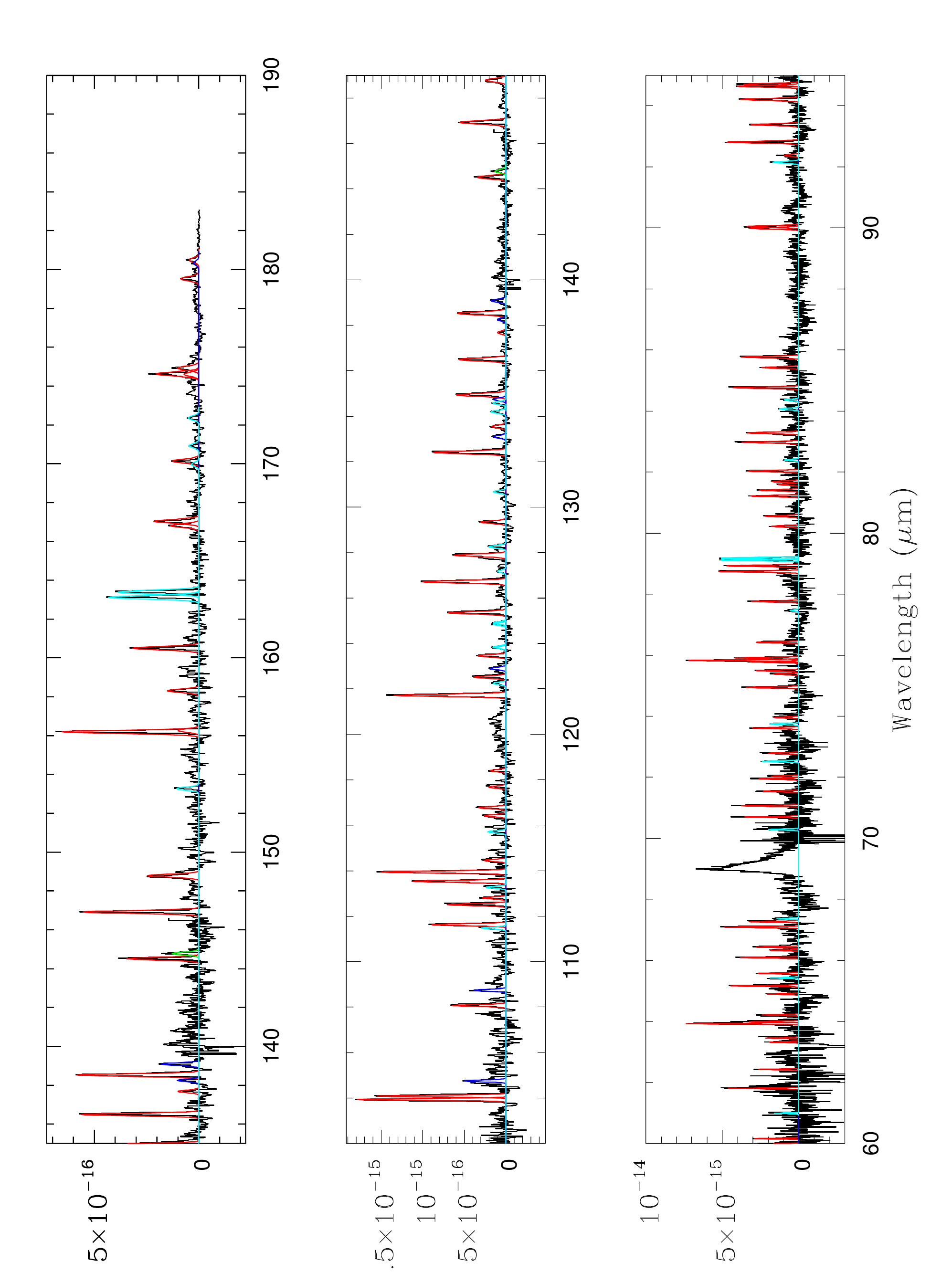}
\caption{The continuum subtracted PACS spectrum of OH~32.8-0.3.}
\end{figure*}

\begin{figure*}
\centering
\includegraphics[width=17cm]{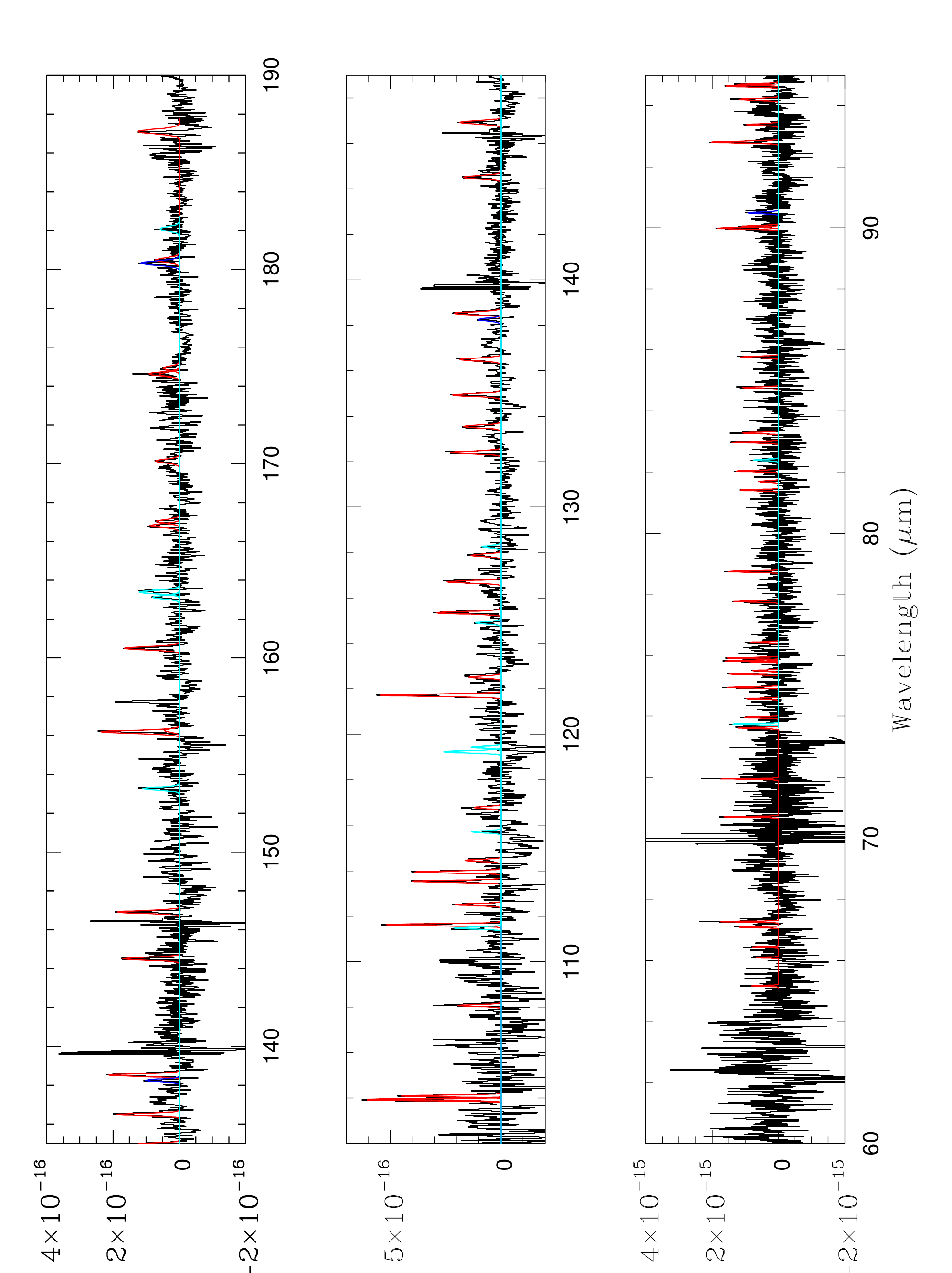}
\caption{The continuum subtracted PACS spectrum of OH~42.3-0.1.}
\end{figure*}

\end{document}